\documentclass[a4paper,11pt]{article}
\pdfoutput=1 

\usepackage{jheppub} 
\usepackage{Definitions}
\usepackage[T1]{fontenc} 
\usepackage{amsfonts,amsmath,amssymb}
\usepackage{bbold}
\usepackage{graphicx}
\usepackage{hyperref}
\usepackage{caption}
\usepackage{physics}
\usepackage{subcaption}
\usepackage{amssymb}
\usepackage{tikz}
\usetikzlibrary { decorations.pathmorphing, decorations.pathreplacing, decorations.shapes}
\usepackage{yfonts}
\usepackage[skins,theorems]{tcolorbox}
\usetikzlibrary {arrows.meta}
\tikzset{cross/.style={cross out, draw=black, minimum size=2*(#1-\pgflinewidth), inner sep=0pt, outer sep=0pt},
cross/.default={3pt}}

\definecolor{blue3}{RGB}{31, 119, 180}
\definecolor{red3}{RGB}{	214, 39, 40}
\definecolor{orange3}{RGB}{255, 127, 14}
\definecolor{green3}{RGB}{44, 160, 44}
\definecolor{repBlue}{RGB}{31, 119, 180}
\definecolor{repRed}{RGB}{	214, 39, 40}
\definecolor{repGreen}{RGB}{44, 160, 44}
\definecolor{violet2}{RGB}{102,0,204}
\definecolor{Blue}{RGB}{214, 39, 40}
\definecolor{Red}{RGB} {31, 119, 180}

\title{\boldmath A non-perturbative construction of the de Sitter late-time boundary}
\author[]{Kamran Salehi Vaziri}

\affiliation[]{Institute of Physics, University of Amsterdam, Amsterdam, 1098 XH, The Netherlands}

\emailAdd{k.salehivaziri@uva.nl}

\abstract{
We propose a new approach for constructing the late-time conformal boundary of quantum field theory in de Sitter spacetime. 
A boundary theory which consists of a continuous family of primary operators residing on unitary irreducible representations, the principal series. 
These boundary operators exhibit two-point functions that include contact terms alongside standard CFT two-point functions. 
We introduce a bulk-to-boundary expansion in which a bulk operator, when pushed to the boundary, is represented as an integral over boundary operators. 
The kernel of this integral is related to the \Kallen spectral density, and we examine the convergence of the expansion by deriving the spectral density’s large dimension limit. 
Additionally, we derive an inversion formula for the bulk-to-boundary expansion, where, given a bulk theory, the boundary operator content is constructed as an integral of the bulk operator times the bulk-to-boundary propagator. 
We verify the inversion formula by recovering the boundary two-point function and reproducing perturbation theory.
Along the way, we define an operator that generates both the bulk-to-boundary and free bulk-to-bulk propagators from the boundary two-point function, proving to be a powerful tool for simplifying de Sitter diagrams. 
}

\begin{document} 
\maketitle
\flushbottom

\newpage
\section{Introduction}\label{sec:intro}
Observations show that we are living in a universe with accelerated expansion. Moreover, it is widely believed that the cosmos began with an exponential expansion, known as the period of inflation~\cite{Guth:1980zm,Linde:1981mu,Baumann:2009ds}. The simplest model to describe both of these epochs of expanding spacetime is de Sitter, a maximally symmetric spacetime with a positive curvature; therefore, it is interesting to study the quantum fields in such a background. 
 The non-perturbative approach to quantum field theory (QFT) in de Sitter where the conformal bootstrap methods can be employed~\cite{Hogervorst:2021uvp,DiPietro:2021sjt}, has proven useful, for example, in dealing with IR divergences~\cite{Gorbenko:2019rza,Anninos:2024fty}.
The late-time boundary of de Sitter is conformal, meaning that the late-time correlation functions satisfy the conformal Ward identities. 
This conformal symmetry is a promising feature of de Sitter that has been exploited widely in the context of perturbation theory~\cite{Maldacena:2002vr,Arkani-Hamed:2015bza,Arkani-Hamed:2018kmz,Lee:2016vti,Sleight:2021plv,DiPietro:2021sjt,Sleight:2019mgd,Goodhew:2020hob,Chen:2017ryl,Salcedo:2022aal,Donath:2024utn}. 
In this work, we approach the late-time boundary of de Sitter from a non-perturbative point of view.

QFT in Anti-de Sitter (AdS) spacetime, the counterpart of de Sitter with a negative cosmological constant, also exhibit a conformal boundary. Moreover, AdS symmetries  provide a simple recipe  for defining the boundary operators through a boundary-operator/bulk-state map~\cite{Paulos:2016fap} which crucially have dimensions matching with the unitary irreducible representations (UIRs) of AdS isometry group. The boundary-operator/bulk-state map also gives rise to a bulk-to-boundary expansion, with finite radius of convergence, when the bulk field is pushed to the boundary at $z= 0$:
\be\label{eq: intro BB AdS}
\phi(z,\V x) = \sum_{i} b_i~ z^{\D_i} \left[\O_i (\V x) +\text{descendants}\right]~,
\ee 
where  we used Poincare coordinates and the sum is over primary operators $\O_i$  with scaling dimension $\D_i$  at boundary point $\V x$. The discrete sum above is reminiscent of the discrete spectrum of AdS -- AdS is a box. 

An attempt to write a bulk-to-boundary expansion similar to~\reef{eq: intro BB AdS} for de Sitter encounters conceptual and technical difficulties. Such expansion would generally produce boundary operators with dimensions that do \textit{not} match de Sitter UIRs -- making their action on a generic state subtle. Consequently, their injection inside correlation functions leads to inconsistencies which make them unnatural objects to bootstrap the boundary correlators.  In this work, we propose an alternative bulk-to-boundary expansion, directly addressing de Sitter features such as its continuous spectrum~\cite{Hogervorst:2021uvp,DiPietro:2021sjt,DiPietro:2023inn}. A bulk-to-boundary expansion that instead of a discrete sum like AdS case, is a continuous sum, an integral over UIRs of de Sitter:
\be
\phi(\eta,\V y) =\int_{\text{UIR}} \, a_\D (-\eta)^\D \Oo_\D(\V y) +\text{descendants}~,
\ee
in which $ \Oo_\D(\V y)$ is a continuous family of primary operators with scaling dimension $\D$ whose correlation functions satisfy the conformal Ward identities. Here we used the planar coordinates~\reef{eq: planar coordinates} with the conformal boundary at $\eta\to0^-$.
In this work, we consider scalar operators and focus on contributions from the principal series representations i.e. $\D=\hd+i\lambda$ with $\lambda\in\Real$.

We find an inversion formula for the bulk-to-boundary expansion, a bulk integral over the bulk operator against a kernel $K_\D$ that turns out to be the bulk-to-boundary propagator:
\be
\Oo_\D(\V{y})= \frac{1}{\ND{\D}} \int_{\text{bulk}} K_{\D}(\eta',\V{y}';\V{y}) \phi (\eta',\V {y}')~,
\ee
where $\ND{\D}$ is a normalization fixed by the \Kallen spectral density.
This inversion formula practically defines the boundary theory: Given a bulk field $\phi$, the inversion formula produces the corresponding boundary operators. The inversion formula also naturally uncovers the existence of boundary contact (local) terms and the non-commutative nature of the boundary operators. 
\subsection*{Outline and summary of results}\label{sec: Outline and summary}
We start by reviewing the basics of de Sitter spacetime such as a few coordinate systems, embedding space formalism, and unitary irreducible representations of de Sitter symmetry group \SOd. We also briefly discuss and set conventions of the bulk-to-boundary propagator and bulk two-point functions. 

In section~\ref{sec:Dhat}, we introduce the \textit{boundary-to-bulk connector}, a differential operator that, when acting on boundary points, lifts them to the bulk and transforms a  boundary two-point function to a bulk-to-boundary propagator and free bulk-to-bulk two-point function. The boundary-to-bulk connector proves to be helpful both at recasting the descendants in bulk-to-boundary expansion as well as simplifying Feynman diagrams in de Sitter -- see appendix~\ref{sec: Identites}. In section~\ref{sec:AdS expansion}, we review the AdS bulk-to-boundary expansion, the AdS boundary-operator/bulk-state correspondence and why a naive AdS-like discrete bulk-to-boundary expansion in de Sitter results in inconsistencies.  

In section~\ref{sec:Bulk-to-Boundary expansion}, we propose the bulk-to-boundary expansion, relate the integral kernel to the \Kallen spectral density and discuss the boundary two-point functions. In section~\ref{sec:Convergence}, we study the convergence of the bulk-to-boundary expansion by deriving the large dimension limit of the \Kallen spectral density and arguing it is dominated by the UV physics. 

In section~\ref{sec:Inversion formula}, we derive the main result of  this paper, the inversion formula. In sections~\ref{sec:proof} and~\ref{sec:Alternative}, we go from the bulk-to-boundary expansion to the inversion formula and back. We verify the inversion formula produces the expected two-point function in a generic theory as well as the perturbation theory in section~\ref{sec:Consistency}. Finally, we discuss the hermitian conjugation of boundary operators and its relation to shadow transform in section~\ref{sec:dagger}.

Several appendices are provided, some of which may be of independent use in other contexts.  In appendix~\ref{sec: large Delta}, we report the large-dimension limit of free propagators in both AdS and de Sitter spacetimes. We also review the role of boundary contact terms in free as well as interacting theories in appendix~\ref{sec: Contact terms}. Additionally, we discuss how boundary operators fail to commute despite being part of a Euclidean theory. In Section~\ref{sec: Identites}, we review and re-derive several essential diagrammatic identities in de Sitter space, including the V-diagram, split representation, and broken-leg identity.
\subsection*{Conventions and notations}\label{sec: notations}
In this work, we take our vacuum to be the Bunch-Davies vacuum and we denote it by $\ket{0}$. We also consider \textit{Wightman} correlation functions in the Bunch-Davies vacuum that are defined by analytical continuation from the sphere~\cite[sec 2.3]{Hogervorst:2021uvp}.
Throughout this paper, we use upper case $X_A$ and $Y_A$ for EAdS and de Sitter points in the embedding space respectively. The upper case Latin indices correspond to the indices in the embedding space $\Mink^{d+1,1}$ with scalar product $\cdot$ defined as the inner product in $\Mink^{d+1,1}$ with mostly plus signature:
\be
Z\cdot Z'= Z^A Z'_A = -Z_0 Z'_0+ Z_1Z'_1+\cdots+Z_{d+1}Z'_{d+1}~.
\ee
We also represent the Euclidean vectors in $\Real^d$ by bold letters, for example, in planar coordinates, the spatial coordinates are represented as $\V y $. The exponents are indicated by lower case Latin letters $y^i$ and the scalar product is indicated by
\be
\V z. \V z'= z^i z'_i = z_1 z_1 +\cdots + z_dz'_d~.
\ee
We also define the square of a vector as $\V z^2 \equiv \V z . \V z$.
We focus on principal series representations and primary operators that have a scaling dimension
\be
\D=\hd+i\lambda~,
\ee
with $\lambda\in \Real$ unless it is stated otherwise. We reserve the letter $\lambda$ for the corresponding scaling dimension throughout the paper. For example, $\D_1 = \hd+i\lambda_1$ and $\D'=\hd+i\lambda'$. We denote the \textit{shadow} of a scaling dimension as $\bar{\D}=d-\D$.

All delta functions in this paper are Dirac delta functions, except where otherwise specified. More precisely:
\be
\delta_{\mu_1,\mu_2} \equiv \delta(\mu_1-\mu_2)~,\qquad   \delta_{\V z_1,\V z_2} \equiv \delta^{(d)} (\V z_1-\V z_2)~.
\ee

We also perform many $d$-dimensional Fourier transformations as well as Hankel transformations. We fixed the conventions regarding these transformations in~\reef{eq: FT notation} and \reef{eq: HT notation}. In several parts of this paper we perform an integral over bulk of de Sitter, where we use the shorthand notations defined as
\be\label{eq: dS volume int}
\int_{\text{dS}} dY = \int_Y \equiv \int_{\eta} \int_{\V y}~\qquad\text{with}\quad  \int_{\eta}\equiv \int_{-\infty}^0 \frac{d \eta}{(-\eta)^{d+1}} \quad,\quad  \int_{\V y}\equiv \int_{\mathbb{R}^d} {d^dy}~.
\ee
When we do not specify the integral region for $d$-dimensional integrals explicitly, that means the integral is over $\mathbb{R}^d$.
In this paper, we consider two types of conformal primary operators: discrete operators, denoted by $\O$, with two-point functions featuring a Kronecker delta of the operator dimensions, and continuous operators, denoted by $\Oo$, whose two-point functions contain a Dirac delta on the dimensions.
\newpage
\section{Preliminaries}\label{sec:Preliminaries}
This section covers the background material used in this paper. We begin by outlining the basics of QFT in rigid de Sitter spacetime, including the Hilbert space, embedding space, and mode functions in de Sitter. We also briefly review the two-point function and the \Kallen spectral decomposition in de Sitter, which later on will be used to study the convergence of the bulk-to-boundary expansion. In Section \ref{sec:Dhat}, we introduce the boundary-to-bulk connector and show how it generates both the bulk-to-boundary and bulk-to-bulk propagators from boundary two-point functions. Lastly, we review the bulk-to-boundary expansion in AdS spacetime and discuss why a naive attempt to replicate it for de Sitter space can lead to complications.
\subsection{QFT in de Sitter}\label{sec:QFT in dS}

\subsubsection*{de Sitter spacetime}
de Sitter is a maximally symmetric spacetime with positive curvature. De Sitter spacetime in $d+1$ dimensions (\dS$_{d+1}$) can be represented as the embedding of the set of points that are a distance $R$ from the origin in Minkowski space $\Mink^{d+1,1}$ (with mostly $+$ signature):
\be\label{eq: Embedding dS}
-Y_0^2+Y_1^2+\cdots+Y_{d+1}^2=R^2~,
\ee
where $R$ is the de Sitter radius. This indicates that \dS$_{d+1}$ isometry group consists of the rotations that leave the hypersurface in~\reef{eq: Embedding dS} invariant, i.e. \SOd~.
On this manifold we can place different slicings and coordinates. Global coordinates that cover all of de Sitter are defined as
\be\label{eq: global coordinates}
Y^0=R \sinh t~, \qquad Y^a=R\, \Omega^a \cosh t~,
\ee
in which $t\in \Real$, $a$ runs over $\{1,\cdots,d+1\}$ and $\Omega^a$ is a unit vector on $S^d$ ($\Omega^a \Omega_a=1$). The global coordinates induce the following metric:
\be\label{eq: global metric}
ds^2=R^2(-dt^2+\cosh^2 t\, d\Omega^2_d)~,
\ee
where $d\Omega^2_d$ represents the unit $S^d$ metric. By a change of variable $\sinh t = \tan \tau$, we find the conformal coordinates with the metric:
\be\label{eq: conformal metric}
ds^2 = R^2 \frac{-d\tau^2+d\Omega^2_d}{\cos^2 \tau}~.
\ee
These coordinates illustrate the fact that de Sitter is conformally equivalent to finite cylinder $(-\pi/2,\pi/2)\times S^d$.

And finally, we introduce the planar coordinates $y^\mu = (\eta,\V y)\in (\Real_{-} \times \Real^d)$ which is the main coordinate system we use in this paper:
\be\label{eq: planar coordinates}
Y^0=R \frac{\eta^2-\V{y}^2-1}{2\eta}~,\qquad Y^i =-R \frac{y^i}{\eta}~,\qquad Y^{d+1}=R \frac{\eta^2-\V{y}^2+1}{2\eta}~,
\ee
in which $i$ runs over $\{1,\cdots,d\}$. Thus, the planar coordinates have the metric:
\be\label{eq: planar metric}
ds^2= R^2\frac{-d\eta^2+d\V{y}^2}{\eta^2}~.
\ee
We refer to $\eta\to-\infty$ as the past infinity and $\eta\to 0^-$ as the future infinity or the late-time boundary. From now on, when we refer to the de Sitter boundary, we simply mean the late-time boundary of de Sitter.
Note that planar coordinates cover half of de Sitter that corresponds to the causal future of an observable placed at the south pole of the global $S^d$. You may also note that with $\eta\to\pm i z$, one finds that the metric~\reef{eq: planar metric} becomes the Euclidean Anti-de Sitter metric in Poincare coordinates. This outlines the basic idea behind the dS-to-EAdS~\cite{Sleight:2020obc,DiPietro:2021sjt,Sleight:2021plv} dictionary and analytical continuation to EAdS~\cite{Bros:1995js,Loparco:2023rug}. From now on, we will set de Sitter radius to unity, $R=1$. 

\subsubsection*{Embedding space}
It is often useful to picture de Sitter as an embedding in Minkowski. In particular, when dealing with the boundary points, this becomes a useful tool. 
We use the dot product for contraction with respect to Minkowski metric. In other words, de Sitter is  $Y^2\equiv Y\cdot Y = Y_AY^A=1$. The only scalar product made out of two points in de Sitter is the \textit{two-point invariant} and is given by: 
\be
\sigma \equiv Y_1 \cdot Y_2 =   \frac{\eta_1^2+\eta^2_2-\V{y}^2_{12}}{2\eta_1\eta_2}~, \qquad \V{y}_{12}\equiv \V{y}_{1}-\V{y}_{2} ~,
\ee
where we spelled out its explicit expression in planar coordinates.
Note that the distance between two-point is related to $\sigma$ via
\be
(Y_1-Y_2)^2=(Y_1-Y_2)\cdot (Y_1-Y_2)= 2-2\sigma~,
\ee
which shows that the spacelike separations correspond to $\sigma<1$, timelike separations to $\sigma>1$, and lightlike separations and short distance singularity correspond to $\sigma=1$.

The late-time boundary $\eta\to 0^-$ is associated with approaching the lightcone passing through the origin in $\Mink^{d+1,1}$. Therefore, it can be identified by the null vector $P^A$ that satisfies~\cite{Costa:2011mg,Costa:2014kfa,Loparco:2023rug}: 
\be
P^2=0~, \qquad P^A\sim\Lambda P^A~,\qquad P^0>0~,
\ee
where $\Lambda\in\Real^+$ and $\sim$ indicates that the point are identified under a real rescaling in Embedding space. In particular, in planar coordinates, the boundary points are parametrized as:
\be\label{eq: P and y}
P^0_{\V{y}}=\half\left(\V y^2+1\right)~, \qquad P^i_{\V{y}}=\V{y}^i~,\qquad P^{d+1}_{\V{y}}= \half\left(\V y^2-1\right)~.
\ee 
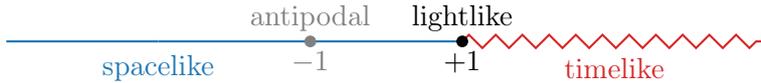
\begin{figure}
\begin{center}
\begin{tikzpicture}
\draw[thick,blue3]                    (-5,0) -- (-3,0)  node[anchor=north]{};
\draw[thick,blue3]                    (-3,0) -- (-1,0) ;
\draw[thick,blue3]                    (-1,0) -- (+1,0);
\filldraw[gray] (-1,0) circle (2pt) node[anchor=north]{$-1$} node[anchor=south]{antipodal};
\draw[decorate,decoration=zigzag,thick,red3](1,0) -- (5,0) node[anchor=west]{};
\filldraw[black] (+1,0) circle (2pt) node[anchor=north]{$+1$} node[anchor=south]{lightlike};
\node[scale=1,red3] at (3,-0.35) {timelike};
\node[scale=1,blue3] at (-3,-0.35) {spacelike};
\end{tikzpicture}
\end{center}
\caption{Range of the two-point invariant $\sigma=Y_1\cdot Y_2$.} \label{fig:sigma}
\end{figure}
\subsubsection*{Hilbert space}
Let us now briefly review the unitary irreducible representations (UIRs) of dS symmetry group. 
In this paper, we focus on scalar operators and UIRs.  For a more detailed and pedagogical discussion on both scalar and spinning representations, see for example~\cite{Sun:2021thf,Penedones:2023uqc,Schaub:2023scu,Letsios:2023qzq,Dobrev:1977qv}. 

The de Sitter symmetry group is the Euclidean conformal group~\SOd~with the conformal algebra generated by dilatation $D$, $SO(d)$-rotation $M_{i j}$, momentum $P_i$ and the special conformal transformation $K_i$. The corresponding Killing vectors of these generators are:
\ba\label{eq:CFT algebra}
D:~&\eta\partial_\eta +y^i \partial_i \qquad & M_{ij}:~& y_j \partial_i-y_i\partial_j\\
P_i:~& \partial_i \qquad & K_i:~&(\eta^2-y^2) \partial_i + 2 y_i \,\eta \partial_\eta+ 2y_i \, y^j\partial_j
\ea  
where $\partial_i \equiv \frac{\partial}{\partial y^i}$ and $ \partial_\eta = \frac{\partial}{\partial \eta}$. Note that  upon moving to the late-time boundary $\eta=0$, the generators become the standard generators of the conformal algebra in Minkowski space.

A (scalar) irreducible infinite dimensional representation of \SOd\, is specified by a complex variable $\D$, often called scaling dimension. The scaling dimension is related to the Casimir eigenvalue $\D(d-\D)$. Scalar \textit{unitary} irreducible representations fall under the following categories:\footnote{For $d=1$, exceptional series coincide with discrete series and they form both highest-weight and lowest-weight representations~\cite{Sun:2021thf,Penedones:2023uqc}.}\footnote{Type II exceptional series appear for spinning irreps.}
\begin{itemize}
\item \textbf{Principal series}, $\CP_\D$: $\D=\hd+i\lambda$ with $\lambda \in \Real$.
\item \textbf{Complementary series}, $\CC_\D$: $0<\D<d$.
\item \textbf{Type I exceptional series}, $\mathcal{V}_{p}$ : $\D=d+p-1$ with $p\in \mathbb{Z}_{>0}$.
\end{itemize}
For example, free massive particles with mass $m^2= \D(d-\D)$ are described by the principal series ($m\geq \hd$) and the complementary series $(m<\hd)$. The shift-symmetric scalars correspond to exceptional series type I~\cite{Sun:2021thf,Penedones:2023uqc}. In this paper, we will focus on principal series. 
It is also worth mentioning that the complementary series and principal series with scaling dimension $\D$ and $\Db=d-\D$ are isomorphic as can be seen by the shadow transformation.  

The Hilbert space of QFT in de Sitter can be represented by states that are labeled by the eigenvalues of simultaneously diagonalizable generators of the algebra, i.e. Cartan subalgebra.  Such a Cartan subalgebra can be formed from the generators of $d$-dimensional translations,  $P_i$, whose eigenvalues we call spatial momentum and denote them by $\V{k}$:
\be\label{eq: k states}
\ket{\D,\V{k}}~: ~\qquad  P_i \ket{\D,\V{k}} = k_i \ket{\D,\V{k}}~.
\ee
These states span the Hilbert space and we denote them by the momentum basis. This basis is delta-function normalizable:
\be\label{eq: k states ortho}
\BBraket{\D',\V{k}'}{\D,\V{k}} = (2\pi)^d \delta_{\V{k},\V{k}'} \delta_{\lambda,\lambda'}~.
\ee
These are very similar to Fock space states used in cosmology literature defined as ${\ketT{\D^*,\V{k}}} \equiv a^\dagger_{\V{k}}\ket{0}$ with the difference that the Fock space states are defined within a specific mutliplet labeled by $\D^*$ which is defined by the mass in free massive scalar $m^2=\D^*(d-\D^*)$. On the other hand, the states $\ket{\D,\V{k}}$ are delta-function normalizable in space of the scaling dimensions, hence they are a good candidate to use in spectral decompositions. In particular, the resolution of the identity is given by~\cite{Hogervorst:2021uvp}:
\be
\mathbb{1} = |0 \rangle \langle 0| + \sum_A \int_{-\infty}^{\infty}\!d\lambda \int\!\frac{d^dk}{(2\pi)^d} \; \ket{\Delta,\V k}_{A} \ ^{A} \bra{\Delta,\V k} ~ + ~\text{other irreps}~,
\ee
where the integral is over principal series, $A$ schematically denotes the sum over spin and the contribution from other representations is implicit.

Another commonly chosen basis for the Hilbert space are the states that are labeled by $d$-dimensional \textit{spatial} vector $\V{y}$ and they are related to states in~\reef{eq: k states} through~\cite{Hogervorst:2021uvp}:
\be\label{eq: k. to y states}
\ket{\D,\V{y}} = \int \frac{d^dk}{(2\pi)^d} e^{i\V{k}\cdot \V{y}} \,k^{i\lambda} \ket{\D,\V{k}}~.
\ee
Again, note that $\D=\hd+i\lambda$. 
These states transform like conformal primary operators with dimension $\D$ under the action of de Sitter isometries. More explicitly:
\ba\label{eq:dS isomoteries on states}
P_i \ket{\D,\V{y}}&= \partial_i \ket{\D,\V{y}}~,\\
D\ket{\D,\V{y}}&=(\V{y}.\partial_\V{y}+\D)\ket{\D,\V{y}}~,\\
K_i\ket{\D,\V{y}}&= \left(2y_i (\V y.\partial_{\V{y}}+\D)-y^2 \partial_i\right)\ket{\D,\V{y}}\\
M_{ij}\ket{\D,\V{y}}&=\left(y_j \partial_i-y_i\partial_j\right)\ket{\D,\V{y}}~.
\ea
For a more extended discussion and the relation of the isometries' action on the momentum basis, see~\cite[Appendix B]{Hogervorst:2021uvp}. Now, using eq.~\reef{eq: k states ortho} and~\reef{eq: k. to y states} one finds that the position basis is Dirac delta normalizable: 
\be
\BBraket{\D',\V y'}{\D,\V y}=\delta_{\lambda,\lambda'} \delta_{\V y,\V y'}~.
\ee
The resolution of the identity is also translated to
\be\label{eq: pos res of identity}
\mathbb{1} = |0 \rangle \langle 0| + \sum_A \int_{-\infty}^{\infty}\!d\lambda \int\!{d^dy} \; \ket{\Delta,\V y}_{A} \ ^{A} \bra{\Delta,\V y} ~ + ~\text{other irreps}~.
\ee
Alternatively, one can label these states with the boundary points $P$ in embedding space using equation~\reef{eq: P and y}:
\be
\ket{\D,P_{\V{y}}} \equiv \ket{\D,\V y}~.
\ee
Notice that these states are living on Cauchy surfaces (constant $\eta$ slices) in the \textit{bulk} and they form a basis for the Hilbert space of the bulk theory. The construction of these states does not rely on any boundary point: We first start with the Cartan subalgebra, introduce the momentum basis, define the position basis using the Fourier-like transformation~\reef{eq: k. to y states} for some vector $\V y$ in $\Real^d$. However, one can assign these states to the boundary points using~\reef{eq: P and y} and see that they behave as local primary operators at $\V{y}$. To avoid clutter, we drop $\V y$ index in $\ket{\D,P_{\V{y}}}$.  An important feature of these states is that they satisfy the homogeneity condition~\cite{Loparco:2023rug,Costa:2014kfa}:
\be\label{eq: homo P}
\ket{\D,\alpha P}=\alpha^{-\D} \ket{\D,P}~,\qquad ~\text{for}\quad \alpha>0~. 
\ee 
The homogeneity condition manifests the equivalence of the states (up to a normalization) after rescaling $P \to \alpha P$ as well as its scaling dimension. This property will prove to be extremely useful, for instance in calculating the mode-functions in~\reef{eq:wave function} or inversion formula~\reef{eq: the inversion formula}.
\subsection{de Sitter mode-function and two-point function}\label{sec:dS two-point}
In this section, we will discuss the wave-function and de Sitter Wightman two-point function. 
From now on we pick our vacuum state to be Bunch-Davies vacuum which is denoted by $\ket{0}$. This choice of vacuum is defined by analytical continuation from sphere, hence the alternative terminology: the Euclidean vacuum.
This choice of vacuum also corresponds to the Minkowski vacuum in early times~\cite{Mukhanov:2007zz}.

We define the de Sitter mode-function or the wave-function as\footnote{Not to be confused by the wave-function of the universe in the literature.}
\be\label{eq:wave function}
\Phi_{\V{y}'}(\eta,\V{y}) \equiv \BBBraket{0}{\phi(\eta,\V{y})}{\D,\V{y}'}~.
\ee
This object is fixed by de Sitter symmetries, as the solution to the Casimir equation (equation of motion), up to a theory-dependent normalization $\textgoth{C}_\D$~\cite{Mukhanov:2007zz,Hogervorst:2021uvp}:
\be\label{eq:wave function K}
\Phi_{\V{y}'}(\eta,\V{y}) = \textgoth{C}_\D \, K_\D(\eta,\V y;\V{y}')~,
\ee
 where 
 \be\label{eq:Bb K}
 K_\D(\eta,\V y;\V{y}')= \frac{(-\eta)^\D}{\left(\left(\V y - \V y'\right)^2-\eta^2\right)^\D}~.
\ee
We will shortly see that $ K_\D(\eta,\V y;\V{y}')$ is simply the bulk-to-boundary propagator up to a normalization. 
Note that for $|\eta|>|\V y - \V y'|$, $ K_\D(\eta,\V y;\V{y}')$ has a branch cut. This is where the bulk point $Y=(\eta,\V y)$ is in causal contact with the boundary point $P_{\V y'}$. For later convenience, let us distinguish being above or below the cut in~\reef{eq:Bb K} by defining:
\be\label{eq: K i eps}
 K^{\pm}_{\D}(\eta,\V{y};\V{y}') =  K_{\D}(\eta\pm i \epsilon,\V{y};\V{y}')~.
\ee

Alternatively, one can use the power of embedding space formalism and derive~\reef{eq:wave function K}. To see this,  first note that  $\BBBraket{0}{\phi(Y)}{\D,P_{\V y'}}$ can be only a function of scalar product $P\cdot Y$. Moreover, the state $\ket{\D,P}$ has a fixed scaling dimension according to the homogeneity condition discussed in~\reef{eq: homo P}. These two facts tell us that the wave-function is fixed up to a normalization:
\be
\BBBraket{0}{\phi(Y)}{\D,P_{\V y'}}={\textgoth{C}_\D}~K_\D(Y,P)~,
\ee  
where
\be
K_\D(Y,P) =\frac{1} {(-2P\cdot Y)^\D}~.
\ee
Now, using the explicit expression of $Y$ and $P$ in planar coordinates from equations~\reef{eq: planar coordinates} and~\reef{eq: P and y}, one finds~\reef{eq:wave function K}.
\subsubsection*{\Kallen and spectral density}
A Wightman two-point function in de Sitter admits the following so-called \Kallen spectral representation~\cite{bros,BROS199122,Hogervorst:2021uvp,Loparco:2023rug,DiPietro:2021sjt}
\be\label{eq:Kallen}
G_\phi(\sigma)=\Braket{\phi(\eta_1,\V{y}_1)\phi(\eta_2,\V{y}_2)}=\int_{-\infty}^\infty d\lambda~ \rho(\D) G^{\text{free}}_\D(\sigma) ~+~ \cdots
\ee
where again $\D=\hd+i\lambda$ and $G^\text{free}(\sigma)$ is the free massive Wightman two-point function given by
\be\label{eq: freeG}
G_\D^\text{free} (\sigma)= \frac{\Gamma(\D)\Gamma(d-\D)}{(4\pi)^\frac{d+1}{2} \Gamma(\frac{d+1}{2})} {}\FF{\D}{d-\D}{\frac{d+1}{2}}{\frac{1+\sigma}{2}}~.
\ee
The unitarity of the bulk theory implies that the spectral density $\rho(\D)$ is a non-negative real number~\cite{Hogervorst:2021uvp,DiPietro:2021sjt,Loparco:2023rug}. Here "$\cdots$" denotes possible contributions from other UIRs. 

We remark that from the shadow symmetry of principal series one finds that $\rho(\D) =\rho(\Db) $. Moreover, as it is readily verifiable the free two-point function is shadow symmetric: $G_\D^\text{free} (\sigma) = G_{\Db}^\text{free} (\sigma)$. Using the Hypergeometric identities, one can split the $G_\D^\text{free} (\sigma)$ two-point function into two parts, a function plus its shadow:

\be\label{eq:psi + psi G}
G_\D^\text{free} (\sigma)= \frak{g}_\D \Pi_\D (\sigma) + \frak{g}_{d-\D}\Pi_{d-\D} (\sigma)~,
\ee
where we define:
\be\label{eq:psi}
\Pi_\D(\sigma)= \frac{1}{\left(2(1-\sigma)\right)^\D} \FF{\D}{\D-\hd+\half}{2\D-d+1}{\frac{2}{1-\sigma}}~,\qquad  \frak{g}_\D =  \frac{\Gamma(\D)\Gamma(\hd-\D)}{4\pi^{\hd+1}}~.
\ee
Note that $\Pi_\D(\sigma)$ is proportional to the EAdS propagator. This again showcases the basic idea behind the dS-EAdS dictionary~\cite{Sleight:2020obc,Sleight:2021plv,DiPietro:2021sjt}.
For later convenience, we use the shadow symmetry of the spectral density and rewrite the \Kallen representation in terms of $\Pi_\D(\sigma)$:
\be\label{eq:KL psi}
G_\phi(\sigma)=\int^\infty_{-\infty}d\lambda~ 2\rho(\D)\,\frak{g}_\D \Pi_\D (\sigma) ~+~ \cdots
\ee

It is interesting to examine what happens to $G_\D^\text{free} (\sigma)$ if we take one of  the points,  say  $Y_2$, to the boundary. In particular, in planar coordinates, this limit is recovered by taking $\eta_2\to0^-$ and one finds:
\ba\label{eq: limit of psi and G}
\Pi_\D(\sigma) &\limu{\eta_2\to0^-} (-\eta_2)^{\D}  K_\D(\eta_1,\V y_1;\V{y}_2)\\
G_\D^\text{free} (\sigma)&\limu{\eta_2\to0^-}  \frak{g}_\D  (-\eta_2)^{\D} K_\D(\eta_1,\V y_1;\V{y}_2)~+~ \D \longleftrightarrow d-\D~.
\ea
As mentioned before, the wave-function is indeed proportional to the bulk-to-boundary propagator. For simplicity, from now on, we will call $K_\D(\eta_1,\V y_1;\V{y}_2)$ the bulk-to-boundary propagator keeping in mind that the limit of the bulk-to-bulk propagator (where we send one point to boundary) is given by sum of  $K_\D(\eta_1,\V y_1;\V{y}_2)$ and its shadow with proper normalization and time scaling factor as illustrated in~\reef{eq: limit of psi and G}.
\subsection{Boundary-to-bulk connector}\label{sec:Dhat}
The late-time boundary of de Sitter is conformal, meaning that its late-time correlation functions  transform as conformal correlation functions under the isometry group of de Sitter and they satisfy the conformal Ward identities~\cite{Arkani-Hamed:2015bza,Arkani-Hamed:2018kmz,Hogervorst:2021uvp,DiPietro:2021sjt}. 
In particular, we expect the existence of boundary primary operators, $\Oo_\D(\V y)$, that satisfy the conformal Ward identities.

Before moving on to de Sitter boundary operators, let us recall some of the basics of CFTs in $\Real^{d}$. The conformal symmetry fixes the structure of the two-point function:
\be\label{eq: CFT twopoint}
\Braket{O_1(\V x_1)O_2(\V x_2)}_{\text{CFT}}= \frac{\delta_{\D_1 \D_2}}{\left|\V x_1 -\V x_2\right|^{2\D_1}}~,
\ee
where $x_{12}=\left|\V x_1 -\V x_2\right|$, $\D_i$ represent the scaling dimension of the primary scalar operator $O_i$, $\delta$ is a Kronecker delta and the operators are normalized to have unit coefficient in the RHS. The unitarity bound implies that $\D\geq \frac{d-2}{2}$.\footnote{Except for the identity operator with $\D=0$.} Note that the two-point function of non-identical operators vanishes due to the Kronecker delta above. 

As we will discuss shortly, the boundary primary operators of a unitary QFT in de Sitter, on the other hand, do not satisfy the unitary bounds. In other words, unitarity in the bulk does not imply unitarity of the boundary theory. Moreover, in addition to the term in~\reef{eq: CFT twopoint}, we expect that de Sitter boundary  two-point functions of principal series operators  to include contact terms which are allowed by conformal Ward identities:
\ba\label{eq: contact inclusion}
\Braket{\Oo_1(\V y_1)\Oo_2(\V y_2)}_{\partial_\text{dS}} &\supset  \frac{\delta_{\D_1 \D_2}}{y_{12}^{2\D_1}}\\
&\supset \delta_{\V y_1,\V y_2}~.
\ea
This phenomenon has been seen in various contexts see for example~\cite{Bousso:2001mw,ArkaniHamed:2015bza,Hogervorst:2021uvp, Anninos:2023lin,Sengor:2021zlc,Anous:2020nxu} and we will discuss it in more detail in section~\ref{sec:Boundary operators} and appendix~\ref{sec: Contact terms}.
\subsubsection{Constructing bulk-to-boundary propagator}

Now that we have briefly reviewed what we expect from the de Sitter boundary two-point function, let us go back to the main focus of this section and find how one would construct the bulk-to-boundary propagator from boundary two-point function. Let us take the bulk-to-boundary propagator, push the bulk point to the boundary and expand it at late-time $\eta_1\to 0$ (or more precisely expand for when $\frac{|\eta_1|}{y_{12}}\ll 1$):
\ba\label{eq: naive BtoB K}
 K_\D(\eta_1,\V y_1;\V{y}_2) \limu{\eta_1\to0^-} \frac{(-\eta_1)^\D}{y_{12}^{2\D}}\left[1+\D \frac{\eta_1^2}{y_{12}^2}+ \D(\D+1)\frac{\eta_1^4}{2y_{12}^4}+\cdots\right]~.
\ea
The Taylor expansion above can be recast as the action of the following differential operator on the CFT two-point function $f(y_{12})=\frac{1}{y_{12}^{2\D}}$:
\be
 K_\D(\eta_1,\V y_1;\V{y}_2) \limu{\eta_1\to0^-} \Dh{}{1} \left[\frac{1}{y_{12}^{2\D}}\right]~,
\ee
where we define
\ba\label{eq:D def}
\Dh{}{}\equiv (-\eta)^\D {}_0F_1(\D-\hd+1,\frac{1}{4} \eta^2 \partial^2_\V{y})~.
\ea
Here $\partial^2_\V{y} = \sum_{i=0}^\infty \partial^2_{y^i}$  is the Euclidean Laplacian and we used the hypergeometric series expansion given by
\be
{}_0F_1(a,z) = \sum_{k=0}^\infty \frac{z^k}{(b)_k k!}~.
\ee  
We call $\Dh{}{}$ the \textit{boundary-to-bulk connector}. It turns out that this operator will also produce the descendants of the primary operators of the late-time boundary with the Laplacian~\cite{Hogervorst:2021uvp}. At this point, $\Dh{}{}$ is simply a compact version of the Taylor expansion in~\reef{eq: naive BtoB K} which is valid for ${|\eta_1|}<{y_{12}}$. 

To generalize the {boundary-to-bulk connector} for the causally connected region ${|\eta_1|}>{y_{12}}$, one needs to add the contribution from the contact term introduced in~\reef{eq: contact inclusion}~. We claim that the full bulk-to-boundary propagator is given by:
\be\label{eq:DtoBulkBoundary}
\tcbhighmath[boxrule=-1pt,arc=3pt,colback=gray!30!white]{
 K^{\pm}_{\D}(\eta_1,\V{y}_1;\V{y}_2) = \Dh{}{1} \frac{1}{y_{12}^{2\D}}\,+\,\xi^{\pm}_{\D}\, \Dhb{}{1}\, \delta_{\V{y}_1,\V{y}_2}
 }
\ee
where
\be\label{eq: xi def}
\xi^{\pm}_{\D}= \frac{\pi^\hd  \Gamma(i\lambda)}{e^{\pm \pi \lambda}\Gamma(\hd+i\lambda)}~.
\ee
where again we used $\D=\hd+i\lambda$. This relation can be seen by the means of Fourier transform. We spell out the details of the derivation in appendix~\ref{sec: D Fourier}. 
Another convenient presentation of the above equation is
\be
 K^{\pm}_{\D}(\eta_1,\V{y}_1;\V{y}_2) = \hat{D}_{\D'}(\eta_1,\V y_1)\mathfrak{H}_{\D,\D'}(\V{y}_1;\V{y}_2)~,
\ee
where we define the generalized boundary two-point structure as
\be
\mathfrak{H}_{\D,\D'}(\V{y}_1;\V{y}_2)= \frac{\delta^\text{K}_{\D,\D'}}{y_{12}^{2\D}}\,+\,\xi^{\pm}_{\D}\, \delta^\text{K}_{\D,\Db'} \delta_{\V{y}_1,\V{y}_2}~,
\ee
in which $\delta^\text{K}$ is the Kronecker delta.
In summary, we now have a recipe to produce the bulk-to-boundary propagator using the generalized boundary two-point functions: the addition of CFT two-point function and the contact term.


\begin{figure}[t!]
   \centering
\scalebox{1}{
\begin{tikzpicture}[line width=1. pt, scale=2]
\draw[lightgray, line width=2.pt] (-1.5,0.75) -- (1.5,0.75);
\draw[black, line width=1.pt] (-1,0.75) -- (1,0.75);
\draw[black, line width=1.pt] (-1,-0.6) -- (1,-0.6);

\draw[lightgray, line width=1.pt]  (-1,0.75) -- (1,-0.6);
\draw[lightgray, line width=1.pt]  (1,0.75) -- (-1,-0.6);

\draw[fill=black] (-1,0.75) circle (.03cm);
\draw[fill=black] (1,0.75) circle (.03cm);
\draw[fill=black] (-1,-0.6) circle (.03cm);
\draw[fill=black] (1,-0.6) circle (.03cm);

\node[scale=1] at (-1,0.9) {$P_{1}$};
\node[scale=1] at (1,0.9) {$P_{2}$};
\node[scale=1] at (-1,-0.75) {$Y_{1}$};
\node[scale=1] at (1,-0.75) {$Y_{2}$};
\node[scale=1,blue3] at (-1.12,0.1) {$\hat{D}$};
\node[scale=1,blue3] at (1.12,0.1) {$\hat{D}$};

\draw[->, dashed, blue3 , line width=0.3pt]  (-1,0.6)   --  (-1,-0.5)   ;
\draw[->, dashed, blue3 ,line width=0.3pt]  (1,0.6)   --  (1,-0.5)   ;
\end{tikzpicture}
}
   \caption{A schematic illustration  of the boundary-to-bulk connector $\hat{D}_\D$ action. It lifts the boundary point to the bulk with the same spatial planar coordinates $\V y$, producing a bulk-to-boundary propagator $K_\D$ (gray lines) out of the generalized boundary two-point structure.}
  \label{fig:D}
\end{figure}
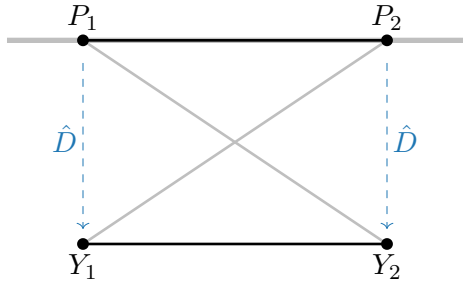

\subsubsection{Constructing bulk-to-bulk propagator}
So far we have seen how operator $\Dh{}{1}$ acting on one point of the boundary lifts that point to the bulk and transform the boundary two-point function to the bulk-to-boundary propagator. One then expects that by acting with $\Dh{}{2}$ on the second point, it would also be lifted to the bulk and produce a bulk two-point function. Following the logic above, one can first expand $\Dh{}{2}$ for $\eta_2\to0^-$ and act with the Laplacian on bulk-to-boundary operator. Then it is straightforward to resum the expansion to find:
\be\label{eq: psi from D on K }
\tcbhighmath[boxrule=-1pt,arc=3pt,colback=gray!30!white]{
\Pi_\D(\sigma)=\Dh{}{2}  K_{\D}(\eta_1,\V{y}_1;\V{y}_2)}
\ee
However, similar to the above, this expansion is valid for small $\eta_2$. One can again make sense of the operation of $\Dh{}{}$ through a Fourier transform. This is done in appendix~\ref{sec: D Fourier} where we verify that the validity domain of~\reef{eq: psi from D on K } can be extended to all values of $\eta_2$. Note that $\Dh{}{2}  K_{\D}(\eta_1,\V{y}_1;\V{y}_2)$ does not fully recover the bulk-to-bulk propagator as it is not shadow symmetric. But, using~\reef{eq:psi + psi G}, after adding the shadow symmetric part with the proper coefficient, one recovers $G_\D^\text{free} (\sigma)$. 

\subsection{AdS bulk-to-boundary expansion and the de Sitter issue}\label{sec:AdS expansion}
Anti-de Sitter space is a maximally symmetric spacetime with an IR cutoff. A QFT in AdS exhibits a discrete spectrum consisting of towers of primaries and their descendants; hence the analogy of AdS to an infinite box. In what follows, we focus on Euclidean Anti-de Sitter. Most of our discussion would translates to Lorentzian AdS through analytical continuation. The AdS analog of the planar coordinates defined in~\reef{eq: planar metric}, is the Poincare coordinates with the metric
\be
ds^2 = R^2 \frac{dz^2 + d\V x^2}{z^2}~,
\ee
where $z>0$, $R$ is the radius of AdS and $\V x$ is a $\Real^d$ vector. $z= 0$ is the conformal boundary where the AdS isometry group acts as the conformal group $\Real^d$. By changing the variables $z=e^T \cos\rho$ and $x= e^T\sin \rho$ one finds the global coordinates with the metric:
\be
ds^2 = R^2\frac{dT^2+d\rho^2 +\sin^2 \rho\, d\Omega^2_{d-1}}{\cos^2\rho}~,
\ee
in which $T\in\Real$ and $0<\rho<\frac{\pi}{2}$. It is easy to see that in these coordinates, AdS is identical to inside of a cylinder with radius $\frac{\pi}{2}$.

\subsubsection*{AdS boundary-operator/bulk-state correspondence}
\begin{figure}[t!]
   \centering
           \begin{tabular}{cc}
\scalebox{1}{
\raisebox{-33pt}{
\begin{tikzpicture}[line width=1. pt, scale=2.5]

\draw[-,blue3!33] (0.75,0) arc (0:180:0.75) -- cycle;
\draw[-,blue3!66] (0.5,0) arc (0:180:0.5) -- cycle;
 \draw[-,blue3] (0.25,0) arc (0:180:0.25) -- cycle;
 
\draw[lightgray, line width=2.pt] (-1,0) -- (1,0);
\node[scale=1] at (1.3,0.025) {$z=0$};
\draw  (0,0)  node[black,fill, circle, scale=0.5]  {};
\end{tikzpicture}
}
}& ~~~~
\scalebox{1}{
\raisebox{-33pt}{
\begin{tikzpicture}[line width=1. pt, scale=2.5]
\draw[lightgray, line width=2.pt] (-1,0) -- (1,0);
\node[scale=1] at (1.3,0.025) {$\eta=0$};
\draw [red3!33, thick,  line width=1pt ,domain=-1:1, smooth ,samples=500] plot (\x, {1-sqrt(\x*\x+1+1)});
\draw [red3!66, thick,  line width=1pt ,domain=-1:1, smooth ,samples=500] plot (\x, {2-sqrt(\x*\x+2^2+1)});
\draw [red3, thick,  line width=1pt ,domain=-1:1, smooth ,samples=500] plot (\x, {6-sqrt(\x*\x+6^2+1)});
\end{tikzpicture}
}
}
\end{tabular}
   \caption{Left: AdS in Poincaré coordinates, with its boundary positioned at $z=0$. The constant global time slices $e^T =\sqrt{z^2+x^2}$, depicted by blue hemispheres, shrink to the boundary origin as one moves backward in time. Right: de Sitter in planar coordinates, where constant global time slices $\sinh t =\frac{\eta^2-y^2-1}{2\eta}$ , shown as red hyperboloids, converge to $\eta=0$ as they approach future infinity.}
  \label{fig:AdS}
\end{figure}
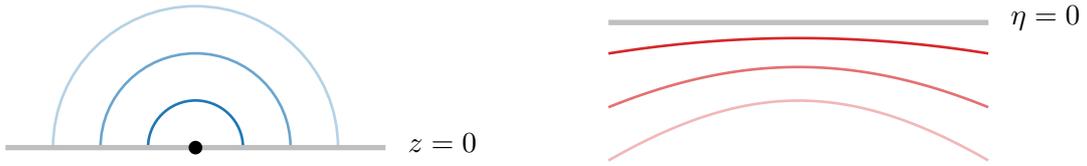

As shown in figure~\ref{fig:AdS}, the constant global $T$ hypersurfaces correspond to hemispheres with radius $\sqrt{z^2+x^2}=e^T$ that end on the boundary $z=0$. In particular, moving back in global time shrinks the hemisphere to the boundary origin $x=0$. This is the main reason behind the Boundary Operator/Bulk State Correspondence~\cite{Paulos:2016fap}: The bulk states that are defined on a constant global T (defined by radial quantization in Poincare coordinates) can be expanded in the eigenbasis of the Hamiltonian $H$ on the cylinder that generates global time translations. The global time translations generated by $H$ correspond to dilatations $D$ around $x=0$ in Poincare coordinates. By evolving (the Hamiltonian eigenstates) back in global time, one shrinks the hemisphere and localizes on the $z=0$ boundary, giving rise to a boundary operator. Similarly, insertion of a boundary operator prepares a state in later time $T$. Therefore AdS exhibits a  boundary-operator/bulk-state map.

In particular, both the states and boundary operators  can be organized into representations of the conformal group \SOd~and labeled by scaling dimensions $\D$ which satisfy the unitary bound. Moreover, the unitarity of a theory either on the boundary or bulk implies the unitarity of the dual theory.  
\subsubsection*{AdS bulk-to-boundary expansion} 
The extrapolate dictionary, tells us that the AdS boundary operators can be defined by pushing the bulk operators to the boundary. More precisely for a bulk field $\phi$ there exists the bulk-to-boundary expansion~\cite{Paulos:2016fap}:
\ba\label{eq: BB AdS expansion}
\phi(z,\V x) &= \sum_{i} b_i~ z^{\D_i} \left[\O_i (\V x) +\text{descendants}\right]\\
&= \sum_{i} b_i~ \hat{\CD}_{\D_i} (z,\V x)~\O_i (\V x) ~,
\ea
in which $b_i$ is called the bulk-to-boundary coefficient, $\D_i$ is the dimension of the conformal primary operator. In the second line, similar to section~\ref{sec:Dhat}, we used the AdS bulk-to-boundary propagator expansion in small $z$ as well as the conformal Ward identities which fix the boundary two-point function to recast the contribution of descendants in a compact form by defining 
\be
\hat{\CD}_{\D_i} (z,\V x)\equiv z^\D~ {}_0F_1(\D-\hd+1,-\frac{1}{4} z^2 \partial^2_\V{x})~.
\ee 
A simple analytical continuation of $\eta\to\pm i z$ on~\reef{eq:D def} would also recover the above expression. 
Due to finite radius of convergence of the AdS  bulk-to-boundary expansion, one can safely insert this expansion inside correlation functions and bootstrap the correlators~\cite{Paulos:2016fap}. 

\subsubsection*{Moving to dS: issues with the naive bulk-to-boundary expansion}
One may use the similarity between the de Sitter planar coordinates and the AdS Poincare coordinates and try to mimic {AdS bulk-to-boundary expansion} in~\reef{eq: BB AdS expansion} for de Sitter to find:
\ba\label{eq: BB dS discrete expansion}
\phi(\eta,\V y) &= \sum_{i} b_i~ (-\eta)^{\D_i} \left[\O_i (\V y) +\text{descendants}\right]~.
\ea
In what follows, we discuss the issues arising from considering such expansion. Notice a direct mimicry of the  AdS bulk-state/boundary operator map is not possible here, as the constant global time hypersurfaces (or any Cauchy surface) do not shrink to a point on the de Sitter boundary.

By inserting the above expansion~\reef{eq: BB dS discrete expansion} in the \Kallen representation~\reef{eq:KL psi}, and taking the late-time limit of the free two-point function (keeping the leading terms) one finds\footnote{In this analysis, we ignore the contact terms to avoid clutter, but they can be restored consistently. We also focus on the cases that only involve the principal series, for a more complete discussion see~\cite{Loparco:2023rug}.}:
\be
\Braket{\phi(\eta,\V y_1)\phi(\eta,\V y_2)}\limu{\eta\to0}  \sum_{ij} (-\eta)^{\D_i+\D_j}b_i b_j \Braket{\O_1(\V y_1)\O_2(\V y_2)} \\
= \int_{-\infty}^{\infty} d\lambda\, 2\rho(\D)\, \frak{g}_\D \frac{(-\eta)^{2\D}}{y_{12}^{2\D}}~.
\ee
Since the boundary operators satisfy the conformal Ward identities, the boundary two-point function takes the form of a CFT two-point function with matching scaling dimensions. With a set of assumptions, including assuming that spectral density is meromorphic -- see~\cite{Hogervorst:2021uvp} for more details, one can deform the contour of RHS and pick up the poles of $\rho(\D)$ to find:
\be\label{eq: KL to boundary}
\Braket{\phi(\eta,\V y_1)\phi(\eta,\V y_2)}\limu{\eta\to0}  \sum_{i} b_i^2 \frac{(-\eta)^{2\D_i}}{y_{12}^{2\D_i}} = \sum_j 2 \lambda_j \frak{g}_{\D_j} \text{Res}[\rho(\D_j)] \frac{(-\eta)^{2\D_j}}{y^{2\D_j}_{12}}~,
\ee
where $\D=\D_j$ correspond to the poles of $\rho(\D)$ in complex $\D$ plane with $\Re [\D]>\hd$. There exist contributions from the spurious poles  $\frak{g}_{\D_j} $ that we did not spell out here but they are expected to be canceled by the zeros of $\rho_\D$~\cite{Hogervorst:2021uvp}. By matching the two sides of the equation above, we conclude that the bulk-to-boundary expansion defined in~\reef{eq: BB dS discrete expansion} implies that the boundary operators are the poles of the \Kallen spectral density, and the bulk-to-boundary coefficients are given by the residues of the \Kallen spectral density. 

Let us clarify the expected boundary operator content for a typical bulk theory using~\reef{eq: KL to boundary}. Consider a massive free theory with a (heavy) field $\phi$ with mass $m_\phi>\hd$ and scaling dimension $\D_\phi$. Its spectral density is trivially a sum of two delta functions which can be written as two simple poles pinching:
\ba
\rho(\D) = \half\left(\delta_{\lambda,\lambda_\phi}+\delta_{\lambda,-\lambda_\phi}\right) &=  \lim_{\epsilon\to0} \frac{\epsilon}{2\pi\left(\epsilon^2+(\lambda^2-\lambda_\phi^2)^2\right)} \\
&= \frac{\e}{2\pi} \left[\frac{1}{\lambda^2-\lambda_\phi^2+i\e}-\frac{1}{\lambda^2-\lambda_\phi^2-i\e}\right]~,
\ea
where the scaling dimension is related to the mass via ${\lambda_\phi}^2+\left(\hd\right)^2 =m^2$.
Now using~\reef{eq: KL to boundary} and picking up the poles on the right side (coming from the first term), one finds the known result that the boundary operators are a set of shadow symmetric operators on the principal series: 
\be\label{eq: free expansion}
\phi(\eta,\V y) \limu{\eta\to 0^-}  (-\eta)^{\D_\phi}\sqrt{\frak{g}_{\D_\phi}} \O_{\phi}(\V y) +  (-\eta)^{\Db_\phi}\sqrt{\frak{g}_{\Db_\phi}}\tilde{\O}_{\phi}(\V y) + \cdots~.
\ee
where  $ \O_{\phi}(\V y) $ and $\tilde{ \O}_{\phi}(\V y) $ are conformal primaries with dimensions $\D_\phi$ and $\Db_\phi=d-\D_\phi$ and $\cdot$ stands for higher order corrections.
This is what is used in perturbation theory to calculate the cosmological correlators~\cite{Arkani-Hamed:2015bza,Arkani-Hamed:2018kmz,Arkani-Hamed:2023kig}. Crucially, these operators have dimensions that coincide with the UIR of isometry group of de Sitter i.e. principal series, in particular $\ket{\O_{\phi}}\equiv\O_{\phi} \ket{0}$ has overlap with UIRs. 

So far everything is fine, but boundary operator of $\phi$ in a free theory is a special case. By turning on interactions~\cite{DiPietro:2021sjt,DiPietro:2023inn} or even considering composite operators in free theory, generally the spectral density has poles off the principal series (and other UIRs) in complex $\D$ plane and hence boundary operators in~\reef{eq: BB dS discrete expansion} would exhibit dimensions that do not match with the scaling dimensions corresponding to UIR of de Sitter isometry group. To be more concrete, consider the example of a composite operator $\phi^2$. The \Kallen spectral density for $\Braket{\phi^2(\eta_1,\V{y}_1)\phi^2(\eta_2,\V{y}_2)}$ includes Gamma functions~\cite{Bros:2009bz,Hogervorst:2021uvp,Loparco:2023rug}:
\be
\rho_{\phi^2} (\D)\supset  \Gamma\left(\frac{2\D_\phi-\D}{2}\right) \Gamma\left(\frac{2d-2\D_\phi-\D}{2}\right)~,
\ee
which imply the existence of poles and consequently boundary operators with dimensions off the principal series:
\be\label{eq:phi2 boundary dim}
\D=
\begin{cases} &2\D_\phi + 2\mathbb{N} \\ 
& 2(d-\D_\phi) +2\mathbb{N}
 \end{cases}~\notin ~\CP~.
\ee

\begin{figure}[t!]
   \centering
\scalebox{1}{
\begin{tikzpicture}[line width=1. pt, scale=2]
\draw[-{Stealth}, line width=0.5pt,black!50, scale=1]  (0,-1)  --  (0,1)  ;
\draw[-{Stealth}, line width=0.5pt,black!50, scale=1]  (-1,0)  --  (5,0)  ;
\draw[dashed, line width=0.5pt,black, scale=1]  (0.75,-1)  --  (0.75,1)  ;

\draw (0.75+0.08,0.3) node[cross,red3] {};
\draw (0.75-0.08,0.3) node[cross,red3] {};

\draw (0.75-0.08,-0.3) node[cross,red3] {};
\draw (0.75+0.08,-0.3) node[cross,red3] {};

\draw (1.5,0.6) node[cross,blue3] {};
\draw (1.5,-0.6) node[cross,blue3] {};
\draw (2.25,0.6) node[cross,blue3] {};
\draw (2.25,-0.6) node[cross,blue3] {};
\draw (3,0.6) node[cross,blue3] {};
\draw (3,-0.6) node[cross,blue3] {};

\draw (2,0) node[cross,black] {};
\draw (2.75,0) node[cross,black] {};
\draw (3.5,0) node[cross,black] {};

\node[scale=1] at (5.3,0){$\text{Re}\,\D$};
\node[scale=1] at (0,1.15){$\text{Im}\,\D$};
\node[scale=1,blue3] at (4,0.6){$\cdots$};
\node[scale=1,blue3] at (4,-0.6){$\cdots$};
\node[scale=1,black] at (4.25,0){$\cdots$};
\end{tikzpicture}
}
   \caption{The pole structure of the \Kallen spectral density of the free field two-point function $\Braket{\phi \phi}$ (in red), the composite operator two-point function $\Braket{\phi^2 \phi^2}$ (in blue) and the  CFT primary two-point function (in black). The poles of the spectral density of $\Braket{\phi \phi}$ are pinching on the principal series (the dashed line: $\text{Re}(\D)=\hd$)  while poles of   $\Braket{\phi^2 \phi^2}$ and CFT spectral densities are off the UIRs of de Sitter. }
  \label{fig:rho}
\end{figure}
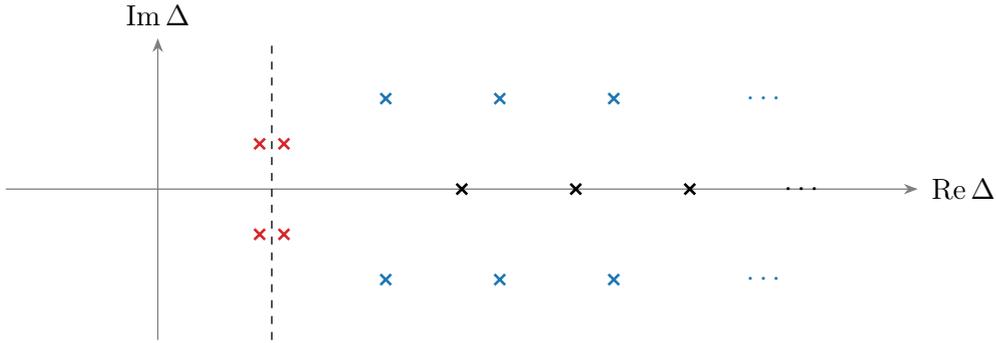

The fact that a generic boundary operator in~\reef{eq: BB dS discrete expansion} does not have a dimension that matches the de Sitter UIR has troublesome consequences. In particular, the insertion of the expansion~\reef{eq: BB dS discrete expansion} in correlation functions would lead to inconsistency and this expansion can only be realized as a formal expansion. Let us illustrate this issue with two examples. Consider the wave-function in~\reef{eq:wave function} for $\phi^2$ and let us push the bulk field to the boundary:
\be\label{eq:wave function phi2}
\BBBraket{0}{\phi^2(\eta,\V{y})}{\D,\V{y}'}\limu{\eta\to0^-}\sum_n b^2_n~ \BBBraket{0}{\O_{n}(\V{y})}{\D,\V{y}'} + \text{descendants}~,
\ee
where the sum is over the conformal primaries of $\phi^2$. Note that $\ket{\D,\V{y}'}$, by definition, satisfies the conformal Ward identities -- see~\reef{eq:dS isomoteries on states}. Using this fact, one concludes that $\BBBraket{0}{\O_{\phi^2}(\V{y})}{\D,\V{y}'}$ would satisfy the same Ward identities as the conformal two-point function $\Braket{O_{\D_n}O_{\D}}$ and in particular such two-point function vanishes when $\D\neq\D_n$. Since the boundary primary dimensions  given by~\reef{eq:phi2 boundary dim} do not match any of UIR of de Sitter group and in particular the principal series, the RHS of \reef{eq:wave function phi2} vanishes leading to a contradiction, since the LHS is indeed a non-zero object. 
Another example is to insert the expansion~\reef{eq: BB dS discrete expansion} into correlation functions, for instance the two-point function, while inserting a complete set of states between the operators:
\ba
\Braket{\phi^2(\eta_1,\V{y}_1)\phi^2(\eta_2,\V{y}_2)} &= \Braket{\phi^2(\eta_1,\V{y}_1)\mathbb{1}\phi^2(\eta_2,\V{y}_2)}\\
&= \int_{-\infty}^{\infty}\!d\lambda \int\!{d^dy} \; \BBBraket{0}{\phi^2(\eta_1,\V{y}_1)}{\Delta,\V y}   \BBBraket{\Delta,\V y}{\phi^2(\eta_2,\V{y}_2)}{0}~.
\ea
Again, using the argument above, the RHS vanishes, leading to a contradiction.
Boundary operators off principal series (and other UIRs of de Sitter group) is not specific to free composite operators, for example, the same issue discussed above arise when one considers a spectral density of the bulk CFT two-point function of primary operators whose poles are on real line: $\D = \Delta_\text{primary}+ \mathbb{N}$~\cite{Hogervorst:2021uvp,Loparco:2023rug}.

Another problem with~\reef{eq: BB dS discrete expansion} is that it is an asymptotic expansion with zero radius of convergence. For instance, for the case of $\phi^2$ two-point function, the bulk-to-boundary coefficients scale as\footnote{This limit can be found by taking the large $\D$ limit of $\rho_{\phi^2} (\D)$ and $\frak{g}_\D$.}
\be
b_n \limu{n\to\infty} n^{\frac{3d-6}{4}}~,
\ee
leading to a non-convergent sum for a finite separation of boundary points $y_{12}$, a small but fixed conformal boundary time and spacetime dimension $d\geq1$.
The contradictions above serve as a reminder that~\reef{eq: BB dS discrete expansion} is a formal expansion and such sum is subtle when it is inserted in correlation function, in particular when the overlap with UIRs of de Sitter is considered. 

\subsubsection*{Possible resolutions}
In summary, the naive de Sitter bulk-to-boundary expansion, inspired by AdS, leads to several issues. 
The boundary operators generically have scaling dimensions that do not belong to unitary irreducible representations of de Sitter group. Hence they are not suitable for injecting the resolution of identity and bootstrapping based on the unitarity of bulk.

Before going on to present the proposed bulk-to-boundary expansion in the next section, let us emphasize that one might come up with various boundary theories that circumvent the issues mentioned above. 
For example,  one might consider a disconnected conformal theory that does not necessarily arise from pushing the bulk operators to the boundary and hence the bulk-to-boundary expansion is not needed in the first place. 
They might as well be unitary CFT with primary operators that satisfy the unitary bound.

Another scenario is to avoid committing to vacuum states and UIRs of  de Sitter bulk theory for boundary operators. In other words, the boundary theory has its own Hilbert space with its own vacuum $\ket{0_\text{boundary}}$ state. Therefore, to make sense of the action of boundary theory on bulk states and in particular UIRs of de Sitter group and the Bunch-Davies vacuum, one needs, for example, to introduce a procedure involving a generalized distributional format to realize the overlap $\BBBraket{0}{\Oo_{n}(\V{y})}{\D,\V{y}'}$. This is similar to taking quasinormal modes as a complete orthogonal basis with respect to an alternative inner product~\cite{Jafferis:2013qia}. 

In what follows, we propose an alternative scenario, involving boundary operators constructed by pushing bulk operators with primary operators which have scaling dimensions on the principal series and hence have non-zero overlap with de Sitter UIRs.

\newpage
\section{Bulk-to-Boundary expansion}\label{sec:Bulk-to-Boundary expansion}
The AdS bulk-to-boundary expansion~\reef{eq: BB AdS expansion} is a \textit{discrete sum} over boundary primary operators. This feature results from the existence of the boundary-operator/bulk-state correspondence as well as the nature of AdS as a box that exhibits a discrete spectrum. 

On the other hand, as shown in the previous section, mimicking~\reef{eq: BB AdS expansion} for de Sitter defines a boundary theory which does not reveal a boundary-operator/bulk-state correspondence and one faces issues upon injecting it into correlation functions. Moreover, the de Sitter spectrum is \textit{continuous}. This has been observed in several cases: \Kallen spectral decomposition of composite operators in free theories such as $\Braket{\phi^2(Y_1)\phi^2(Y_2)}$, $\Braket{\phi\nabla_\mu\phi(Y_1) \phi\nabla_\nu\phi(Y_2)}$, $\Braket{\phi V_\mu(Y_1) \phi V_\nu(Y_2)}$~\cite{Hogervorst:2021uvp,Loparco:2023rug} and stress tensor $\Braket{T^{\mu \nu} (Y_1)T^{\rho \sigma} (Y_2)}$~\cite{Loparco:2024ibp} as well as weakly interacting theories like $\lambda \phi^3$ and $\lambda \phi^4$~\cite{DiPietro:2021sjt,Loparco:2023rug} and strongly interacting theories like the $O(N)$ model~\cite{DiPietro:2023inn} and  CFTs~\cite{Hogervorst:2021uvp,Loparco:2023rug}, all of which have a continuous gapless spectrum on the principal series. The footprint of this fact also exists in study of the tensor products of one particle states as well as the decomposition of bulk CFT primary states into de Sitter UIRs~\cite{Repka:1978,Dobrev:1976vr,Dobrev:1977qv,Penedones:2023uqc}. This is a simple outcome of de Sitter physics: de Sitter generically produces a continuous family of states with mass above the Hubble scale.  

Here, we propose an alternative bulk-to-boundary expansion made for de Sitter, a continuous sum (an integral) of operators that live on de Sitter unitary irreducible representations i.e. the principal series. We propose that by pushing the bulk field to the late-time boundary at $\eta\to0^-$, one can expand the bulk field  in terms of boundary local operators $\Oo (\V y)$ as
\be\label{eq:the expansion}
\tcbhighmath[boxrule=-1pt,arc=3pt,colback=gray!30!white]{
\phi(\eta,\V y) = \int_{-\infty}^\infty d\lambda \, a_\D \Dh{}{} \Oo_\D(\V y)}
\ee
where the integral is over the principal series, $\D=\hd+i\lambda$, the bulk-to-boundary connector $\Dh{}{}$ is defined in~\reef{eq:D def} and $a_\D$ is called the bulk-to-boundary coefficient. Note that in this work, we focus on contributions only from the principal series and in general there can be additional contributions from other representations on the RHS of~\reef{eq:the expansion}. As we will discuss later in more detail, when considering the in-in correlator of the bulk fields, we assume an $i\e$ prescription for  the bulk filed based on its position on the Keldysh-Schwinger contour. The equation above will also inherit the same  $i\e$ prescription i.e. $\eta \to \eta \pm i\e$.

One can easily verify that the action of the conformal generators from~\reef{eq:CFT algebra} on the boundary imply that $\Oo_\D$ satisfy the Ward identities of the conformal primary operators with dimension $\D$. 
The role of $\Dh{}{}$ is to produce the descendants. One can easily see this by expanding $\Dh{}{}$ in small $\eta$ and noticing that the higher order terms in $\eta$ come with derivatives $\partial^2_{\V y}$ which acting on $\Oo(\V y)$, produce the descendants.

In what comes, we will discuss the two-point function of the boundary operators and we will connect  the bulk-to-boundary coefficient to the \Kallen spectral density.   
\subsection{Boundary operators}\label{sec:Boundary operators}
The conformal two-point function in~\reef{eq: CFT twopoint} is not the only solution to the conformal Ward identities. Another solution involves a contact term (a Dirac delta function in position space) that appears when the dimensions of the operators add up to the spacetime dimension $d$. This indeed can occur for operators in the principal series when their dimensions are shadows of each other, i.e., $\D_1 = \Db_2 = d - \D_2$. By definition, the boundary operators $\Oo_\D(\V y)$ reside in the principal series and such contact terms are allowed by the conformal symmetry. Furthermore, as discussed in detail in appendix~\ref{sec: Contact terms}, the contact terms emerge when the free bulk two-point function is pushed to the boundary. Additionally, unitarity also requires the presence of such terms~\cite{Hogervorst:2021uvp}. From these considerations, we expect that a generic de Sitter boundary two-point function takes the form:
\be\label{eq: boundary two-point}
\Braket{\Oo_{\D_1}(\V y_1)\Oo_{\D_2}(\V y_2)} = \alpha_{\D_1}\frac{\delta_{\lambda_1\lambda_2}}{y_{12}^{2\D_1}} +\beta_{\D_1} \delta_{\lambda_1,-\lambda_2} \delta_{\V y_1,\V y_2}~,
\ee
where $\alpha_\D$ and $\beta_\D$ are $\D$-dependent normalization constants that will be determined shortly.
Note that the delta functions above are all Dirac delta functions, as a reminder that the boundary operators are defined as a continuous family.
We remark that in the bulk-to-boundary expansion in~\reef{eq:the expansion}, changing the normalization of the boundary operators by rescaling them is equivalent to changing the bulk-to-boundary coefficient. Here, similar to CFT literature, we pick the normalization of the boundary operators such that the coefficient of the power-law part of RHS of~\reef{eq: boundary two-point} be unity:
\be\label{eq: boundary two-point unity}
\Braket{\Oo_{\D_1}(\V y_1)\Oo_{\D_2}(\V y_2)} =\frac{\delta_{\lambda_1\lambda_2}}{y_{12}^{2\D_1}} +\gamma_{\D_1} \delta_{\lambda_1,-\lambda_2} \delta_{\V y_1,\V y_2}~.
\ee
where $\gamma_{\D}$ is given by~\reef{eq: gamma def}.

A distinctive feature of the de Sitter late-time boundary, despite appearing to be a Euclidean theory, is that its operators do not commute. Specifically:
\ba\label{eq: boundary commutator}
\Braket{[\Oo_{\D_1}(\V y_1),\Oo_{\D_2}(\V y_2)]} &= \left(\gamma_{\D_1}-\gamma_{\Db_1}\right)\delta_{\lambda_1,-\lambda_2} \delta_{\V y_1,\V y_2}~.
\ea
In appendix~\ref{sec: Contact terms}, we explicitly check this for free theory and find:
\be
\Braket{[\Oo_{\D_1}(\V y_1),\Oo_{\D_2}(\V y_2)]} =\frac{1}{2\lambda |\frak{g}_\D|}\delta_{\lambda_1,-\lambda_2} \delta_{\V y_1,\V y_2}~,
\ee
where $\frak{g}$'s absolute value on principal series is also given by $|\frak{g}_\D| = \sqrt{ \frak{g}_\D \frak{g}_{\Db}} $.

The non-commutative feature of de Sitter Euclidean boundary is reminiscent of the Lorentzian nature of the bulk theory. An intuitive way to understand this is by taking boundary operators as the limit of bulk operators being pushed to the boundary. Take two bulk operators, $\phi(\eta_1, \V y_1)$ and $\phi(\eta_2, \V y_2)$. These operators do not necessarily commute when they are in causal contact: 
\be 
[\phi(\eta_1, \V y_1), \phi(\eta_2, \V y_2)] \neq 0 \qquad \text{when} \quad y_{12} < |\eta_1 - \eta_2|~. 
\ee 

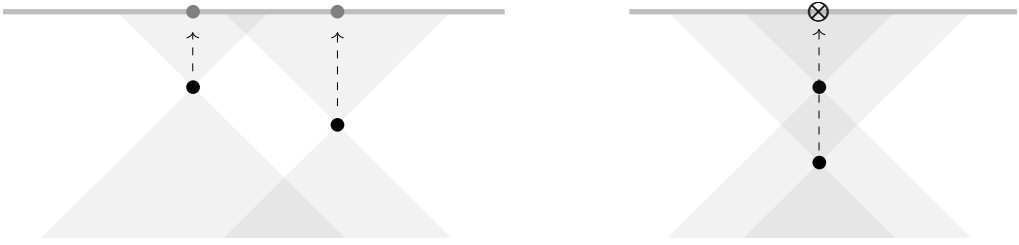
\begin{figure}[t!]
   \centering
           \begin{tabular}{cc}
\scalebox{1}{
\raisebox{-33pt}{
\begin{tikzpicture}[line width=1. pt, scale=2]

\fill[gray,opacity=0.1] (-0.25,0)   -- (-0.25-1,-1) -- (-0.25+1,-1)-- cycle;
\fill[gray,opacity=0.1] (-0.25,0)   -- (-0.25-0.5,0.5) -- (-0.25+0.5,0.5) -- cycle;
\draw  (-0.25,0)  node[black,fill, circle, scale=0.5]  {};

\fill[gray,opacity=0.1] (0.2+0.5,-0.25)   -- (0.2+0.5-1.5/2,-1) -- (0.2+0.5+1.5/2,-1)-- cycle;
\fill[gray,opacity=0.1] (0.2+0.5,-0.25)   -- (0.2+0.5-0.5-0.25,0.5) -- (0.2+0.5+0.5+0.25,0.5) -- cycle;
\draw  (0.2+0.5,-0.25)  node[black,fill, circle, scale=0.5]  {};

\draw[lightgray, line width=2.pt] (0.5-2,0.5) -- (0.5+1.3,0.5);

\draw[->, ,dashed, line width=0.3pt, scale=0.7]  (-0.36,0.15)  --  (-0.36,0.52)  ;
\draw[->, dashed, line width=0.3pt, scale=0.7]  (1.35-0.35,-0.18)  --  (1.35-0.35,0.52)  ;
\draw  (-0.25,0.5)   node[gray,fill, circle, scale=0.5]  {};
\draw  (0.7,0.5)   node[gray,fill, circle, scale=0.5]  {};

\end{tikzpicture}
}
}& ~~~~
\scalebox{1}{
\raisebox{-33pt}{
\begin{tikzpicture}[line width=1. pt, scale=2]

\fill[gray,opacity=0.1] (-0.25,0)   -- (-0.25-1,-1) -- (-0.25+1,-1)-- cycle;
\fill[gray,opacity=0.1] (-0.25,0)   -- (-0.25-0.5,0.5) -- (-0.25+0.5,0.5) -- cycle;
\draw  (-0.25,0)  node[black,fill, circle, scale=0.5]  {};

\fill[gray,opacity=0.1] (-0.25,-0.5)   -- (-0.25-1/2,-1) -- (-0.25+1/2,-1)-- cycle;
\fill[gray,opacity=0.1] (-0.25,-0.5)   -- (-0.25-0.5-0.5,0.5) -- (-0.25+0.5+0.5,0.5) -- cycle;
\draw (-0.25,-0.5)   node[black,fill, circle, scale=0.5]  {};

\draw[lightgray, line width=2.pt] (0.5-2,0.5) -- (0.25+0.8,0.5);

\draw[->, ,dashed, line width=0.3pt, scale=0.7]  (-0.36,0)  --  (-0.36,0.55)  ;
\draw[-, ,dashed, line width=0.3pt, scale=0.7]  (-0.36,0.25-1)  --  (-0.36,0)  ;
\node[scale=1] at (-0.25,0.5) {$\pmb{\otimes}$};

\end{tikzpicture}
}
}
\end{tabular}
   \caption{A cartoon picture of the  non-commutative nature of de Sitter boundary operators. 
   Left: Two bulk points moving to distinct points on the boundary will be causally disconnected, and hence their corresponding boundary operators commute.
   Right: Two bulk points pushed to the same boundary point will asymptotically remain in causal contact, a sign of non-commutative boundary operators.}
  \label{fig:causal}
\end{figure}

In other words, as shown in figure~\ref{fig:causal}, the operators will not necessarily commute if their spatial separation is smaller than their conformal time separation: $y_{12} = |\V y_1 - \V y_2| < |\eta_1 - \eta_2|$. Therefore, when the operators are pushed to the boundary ($\eta_1, \eta_2 \to 0^-$), the commutator between two operators whose spatial separation is $y_{12} = \alpha\, |\eta_1 - \eta_2|$, with $\alpha < 1$ (corresponding to coincident points at $\eta_1,\eta_2\to0$), becomes more subtle and eventually give rise to the contact terms. 
In appendix~\ref{sec: Contact terms}, we will see this non-commutative  feature will come about naturally by considering the canonical commutation relation in the case of the free theory.

\subsection{Bulk-to-boundary coefficient}\label{sec:Bulk-to-boundary coefficient}
The bulk-to-boundary coefficient $a_\D$ is related to the \Kallen spectral density. To see this, one can insert equation~\reef{eq:the expansion} into the bulk two-point function:
\be
\Braket{\phi(\eta,\V y_1) \phi(\eta,\V y_2)} = \int_{-\infty}^\infty d\lambda_1d\lambda_2 \,a_{\D_1}a_{\D_2}\,\hat{D}^1_{\D_1} \hat{D}^2_{\D_2} \;\Braket{ \Oo_{\D_1}(\V y_1) \Oo_{\D_2}(\V y_2)}~.
\ee
where we introduce the short notation $\hat{D}^i_{\D} \equiv  \Dh{}{i}$ to avoid clutter and we focused on equal time bulk two-point function $\eta_1=\eta_2=\eta$. Using the explicit expression of the boundary two-point function introduced in~\reef{eq: boundary two-point}, we can rearrange the equation above to find:
\be\label{eq: expansion into bulk prop}
\Braket{\phi(\eta,\V y_1) \phi(\eta,\V y_2)} = \int_{-\infty}^\infty d\lambda \left[a^2_{\D} \alpha_{\D}\, \hat{D}^1_{\D}\hat{D}^2_{\D} \; \frac{1}{y_{12}^{2\D}} + a_{\D}a_{\Db} \beta_{\D}\, \hat{D}^1_{\D}\hat{D}^2_{\Db}\delta_{\V y_1,\V y_2} \right]~.
\ee

Let us now use the \Kallen spectral decomposition in~\reef{eq:Kallen} and push the bulk point to the boundary. In particular, we must find the late-time behavior of the free bulk two-point function. In appendix~\ref{sec: Contact terms}, we discuss this limit in detail. We take the Fourier transform, move the points to the boundary, and then apply the inverse Fourier transform. We observe the existence of the contact terms through this procedure. The result can be found in~\reef{eq: late free G position}. The  \Kallen  decomposition at the late-time limit takes the form
\be\label{eq: Kallen late}
\Braket{\phi(\eta,\V y_1) \phi(\eta,\V y_2)}  \limu{\eta\to 0^-} \int_{-\infty}^\infty d\lambda\; 2\rho(\D) \left(\frak{g}_\D\,\frac{ (-\eta)^{2\D}}{y_{12}^{2\D}} 
+\frac{\coth(\pi \lambda)}{\lambda} (-\eta)^d\delta_{\V y_1,\V y_2}\right)~,
\ee
where we used the shadow symmetry of $\rho(\D)$ to simplify the final answer. Now one can use the small $\eta$ limit of the boundary-to-bulk connector to match the first terms in~\reef{eq: expansion into bulk prop} and~\reef{eq: Kallen late} to find a relation between the bulk-to-boundary coefficient, $\alpha$ and the  spectral density: 
\be\label{eq: a and rho relation}
a_\D^2 \alpha_\D = 2 \frak{g}_\D \rho(\D)~,
\ee
which exactly matches with what we find later in equation~\reef{alpha and beta from inversion} from the inversion formula inserted into \Kallen decomposition.
Returning to~\reef{eq: expansion into bulk prop}, one can use the action of the boundary-to-bulk connector in equations~\reef{eq:DtoBulkBoundary} and~\reef{eq: psi from D on K } and relate the equation directly to alternative representation of \Kallen in~\reef{eq:KL psi}. Then one finds that $\beta_\D$ and $\alpha_\D$ are not independent; hence, $\beta_\D$ is also related to  the bulk-to-boundary coefficient and the spectral density:
\be\label{eq: beta and rho relation}
\frac{\beta_\D a_{\Db}}{\alpha_\D a_{\D}} =\xi_\D^\mp \qquad \longrightarrow \qquad \beta_\D =e^{\pm \pi \lambda}  \frac{\Gamma(i\lambda)\Gamma(-i\lambda)}{2\pi} \frac{\rho(\D)}{a_\D a_{\Db}}~.
\ee
where the $\pm$ arises from the choice of the $i\e$ prescription in bulk-to-boundary propagator or equivalently boundary-to-bulk connector. Notice since we are considering the \Kallen decomposition of the Wightman function, the $i\e$ prescription of the first and second bulk fileds are opposite of each other. 

It is customary to work with  CFT operators that have unit-normalized two-point function. We would also like to work with operators that have such normalization. In other words, we set $\alpha_\D=1$. This normalization scheme fixes $a_\D$ in terms of \Kallen spectral density through~\reef{eq: a and rho relation} and consequently $\beta$ through~\reef{eq: beta and rho relation}:
\ba\label{eq: gamma def}
\alpha_\D=1,\quad \Longrightarrow \quad & a_\D=\sqrt{2 \frak{g}_\D \rho(\D)}\\
&\gamma_\D\equiv\beta_\D\big|_{\alpha_\D=1} = e^{\pm \pi \lambda} \pi^\hd \sqrt{\frac{\Gamma(i\lambda)\Gamma(-i\lambda)}{\Gamma(\hd+i\lambda)\Gamma(\hd-i\lambda) }}~.
\ea
Notice that in this normalization, the contact term factor is real and positive: i.e. $\gamma_\D>0$. 

\subsection{Convergence}\label{sec:Convergence}
The AdS bulk-state/boundary-operator correspondence implies the convergence of the AdS bulk-to-boundary expansion~\reef{eq: BB AdS expansion} with a finite radius of convergence inside a correlation function. The argument for this convergence is very similar to the one in CFT where no assumptions are required about the density of states or OPE coefficients. In the case of de Sitter, on the other hand, we do not have a similar construction that leads to a bulk-state/boundary-operator correspondence, see figure~\ref{fig:AdS}. 

Let us consider inserting the bulk-to-boundary expansion in~\reef{eq:the expansion} into a correlation function where we also insert the complete set of states on both sides of the expanded operator: 
\ba
&\Braket{\phi(\eta_1,\V y_1)\cdots\phi(\eta_m,\V y_m)\cdots\phi(\eta_n,\V y_n) } \\
&\qquad\qquad =\sum_{i,j}\int d\lambda\, a_{\D} \Dh{}{m} \Braket{\phi(\eta_1,\V y_1)\cdots \ket{\psi_i}\bra{\psi_i}\Oo_\D(\V y_m)\ket{\psi_j}\bra{\psi_j}\cdots\phi(\eta_n,\V y_n) } 
\ea
where the sum above is schematic and could be either a discrete sum or an integral over a continuous spectrum and we inserted the resolution of identity $\mathbb{1} = \sum_i \ket{\psi_i}\bra{\psi_i}$.  We might as well, pick theses states to be the UIR of the de Sitter group in~\reef{eq: pos res of identity}, then $\bra{\psi_i}\Oo_\D(\V y_i)\ket{\psi_j}$ is fixed by symmetry up to a constant analogous to the OPE coefficient. The convergence of the integral above is dependent on the asymptotic behavior of  $\bra{\psi_i}\Oo_\D(\V y_i)\ket{\psi_j}$ and $a_\D$ at large $\D$ limit. 

In what follows, we find the large  $\D$ limit of $a_\D$. We do so by use of the equation~\reef{eq: gamma def} and finding the large $\D$ behavior of $\rho_\D$ and $\frak{g}_\D$. The large $\D$ limit of $\frak{g}_\D$ is  given simply by the Stirling's  formula~\reef{eq: Stirling's}:
\be\label{eq: g limit}
\lim_{\lambda\to \pm\infty} \frak{g}_\D =e^{\pm\frac{i \pi d}{4}} \frac{|\lambda|^{\hd-1}}{2\pi^\hd} e^{-\pi |\lambda|}~,
\ee
where $\lambda \in\Real$. 
\subsubsection{Large $\D$ limit of \Kallen spectral density}\label{sec:large Delta Kallen spectral density}
We may now proceed to calculate the large scaling dimension behavior of the spectral density through the \Kallen inversion formulas. There are two equivalent versions of the inversion formula for the \Kallen spectral density. One version is derived through the analytic continuation to Euclidean Anti-de Sitter (also known as the de Sitter Euclidean inversion formula) given by~\cite{Bros:1995js,Loparco:2023rug}:
\be\label{eq:rhoAdS}
\rho(\D) =  \frac{2\pi^{\frac{d+1}{2}}}{\Gamma(\D)\Gamma(d-\D)\Gamma(\frac{d+1}{2})} \int^{-1}_{-\infty} d\sigma\; (\sigma^2-1)^\frac{d-1}{2} {}_2F_1\left(\D,d-\D,\frac{d+1}{2},\frac{1+\sigma}{2}\right) G(\sigma)~,
\ee
where $G(\sigma)$ is the two-point function of a generic theory. Another version of the inversion formula is derived through the analytic continuation to the sphere (also referred to as the  de Sitter Lorentzian inversion formula) and is given by~\cite{Hogervorst:2021uvp}:
\be\label{eq:RhoFormulaNew}
{\rho(\D)=\beta(\D) \int_1^\infty d\sigma \;  \FF{1-\D}{1-d+\D}{\frac{3-d}{2}}{\frac{1-\sigma}{2}} \text{Disc}[G(\sigma)]}~,
\ee 
where the discontinuity is defined as $ \text{Disc} [f(z)] =f(z+i\e)- f(z-i\e)$ and
\be
\beta(\D) = \frac{-i (4\pi)^\frac{d-1}{2}}{\Gamma\left(\frac{3-d}{2}\right)} \frac{\Gamma(1-\D)\Gamma(1-d+\D)}{\Gamma(\D-\hd)\Gamma(\hd-\D)}~.
\ee
These two formulas are equivalent through the change of the contour from the discontinuity of the two-point function to discontinuity of the hypergeometric~\cite{Hogervorst:2021uvp}. Notice that in~\reef{eq:rhoAdS} integrates over a part of spacelike separation $\sigma<-1$ while~\reef{eq:RhoFormulaNew}  includes the full timelike separation discontinuity. 

To find the large $\D$ limit of the spectral density one needs to take the inversion formula and perform a saddle point approximation. It turns out the relevant inversion formula to use is the sphere inversion formula~\reef{eq:RhoFormulaNew}. An intuitive way to see this is that, the large $\lambda$ limit corresponds to the UV. Therefore the main contribution should be dominated by the behavior of the two-point function at the short distances, $\sigma\to1$ and only the sphere inversion formula covers this range. A direct calculation also shows that the EAdS inversion formula has no saddle point in the range of integration. 

Let us now focus on the large $\D$ limit of the sphere inversion formula. 
For later convenience, with the use of a hypergeometric identity, we split the kernel in~\reef{eq:RhoFormulaNew} into a term plus its shadow:
\be
d\sigma \, \beta(\Delta)~ \FF{1-\D}{1-d+\D}{\frac{3-d}{2}}{\frac{1-\sigma}{2}} = dx\, \left[\Xi_\D(x) +  \Xi_{d-\D}(x)\right]~,
\ee 
with 
\be
\Xi_\D (x) = -i 2^{2\Delta-1} \pi^{\hd-1} \frac{\Gamma(1-\D)}{\Gamma(\hd-\D)}x^{\D-1} \FF{1-\D}{\half+\hd-2\D}{1+d-2\D}{-\frac{1}{x}}~,
\ee
where $x=\frac{\sigma-1}{2}$. In other words the inversion formula takes the form:
\be\label{eq:rho from Xi}
\rho(\D)= \int_0^\infty dx\, \Xi_\D (x) \text{Disc}[G(x)] + \D \leftrightarrow d-\D~.
\ee
Using the integral representation of the hypergeometric function in~\reef{eq: 2F1 int rep}, for a fixed $x$, $\Xi_\D (x)$ in large $\Delta$ limit is given by
\be\label{eq:integral rep of Xi}
\Xi_\D (x) \limu{|\Delta|\to\infty}- i 2^{d-1} \pi^{\frac{d-3}{2}}(-\D)^\frac{3-d}{2} x^\frac{d-1}{2} \int_0^1 dt\, \frac{(1-t)^{d-1}}{(t+x)^\frac{d+1}{2}}\left(\frac{t+x}{t(1-t)}\right)^\D~.
\ee 
In large $\Delta$ limit, the integral is dominated by the saddle point:
\be
t^* = \sqrt{x(x+1)}-x~,
\ee
which leads to the following large $\Delta$ approximation of $\Xi_\D (x)$:
\be
\Xi_\D (x)  \limu{|\Delta|\to\infty}- i (4\pi)^{\frac{d-1}{2}} (-\D)^{1-\hd} \frac{\left((x+1)\sqrt{x}-x\sqrt{x+1}\right)^d}{\left(x(x+1)\right)^\frac{d+2}{4}}e^{-\frac{\Delta}{2} \Theta(x)}~,
\ee
with 
\be
 \Theta(x)= \log\left[1+4 \sqrt{x(x+1)}(2\sqrt{x(x+1)}-1-2x) \right]~,
\ee
which is a real, nonpositive, monotonic function for $x\in \Real^+$ with a zero at $x=0$.

 For the moment, let us focus on $\text{Re}[\D]<c$ with a finite $c>0$. This is in particular true for $\D\in \CP$. We will come back to generic $\D$ later.
The integration over $x$ will be suppressed due to exponential decay (or high oscillation) of $e^{-\frac{\Delta}{2} \Theta(x)}$ term for $|\Theta(x)|\gtrsim1/|\Delta|$.
The main contribution to the large $\Delta$ limit in integral of~\reef{eq:rho from Xi}  can be calculated considering the expansion:
\be
 \Theta(x) \limu{x\to 0} -4\sqrt{x} + O(x^\frac{3}{2})~.
\ee
In other words, the integral is dominated by $\nu\lesssim1$ where we define:
\be
\nu \equiv 2 \sqrt{x} \Delta~.
\ee
Since we are probing the vicinity of $x=0$ in the integral~\reef{eq:integral rep of Xi}, the dominant part of the integral is from $t\sim 0$, hence we choose the change of variable $t=r/\lambda$. 
The relevant limit is therefore:
\be
\Xi_\D \left(\frac{\nu^2}{4\Delta^2}\right) \limu{|\D|\to\infty} \frac{- i (\pi \nu^2)^\frac{d-1}{2}}{\pi (-\D)^{d-2}} \int_0^\infty dr\,\, \frac{e^{-r-\frac{\nu^2}{4r}}}{r^\frac{d+1}{2}}=   \frac{-i2^\frac{d+1}{2}(\pi \nu)^\frac{d-1}{2}}{\pi(-\D)^{d-2}} K_{\frac{d-1}{2}}(\nu)~.
\ee 
Although we start with the assumption that $\text{Re}[\D]<c$, by means of analytic continuation the equation above is true for any $\D$ in complex plane except for around the real line.  

The discussion above shows that the inversion formula in  large $\D$ limit is dominated by two-point function behavior  at $x\to 0$ or equivalently $\sigma\to 1$ that corresponds to short distance limit or in other words UV physics. We assume that our theory is UV complete and becomes a CFT in the UV. This implies that the two-point function of a scalar operator in UV as well as its discontinuity $(\text{for} ~\sigma>1)$ would be given by~\cite{Loparco:2023rug}:
\be
G(\sigma) = \frac{c}{(1-\sigma)^\delta},\qquad \text{Disc}[G(\sigma)] =2i \sin(\pi\delta)  \frac{c}{(\sigma-1)^\delta}~,
\ee 
for some normalization factor $c$. 
Plugging this back into the inversion formula, one finds
\ba\label{eq:}
\rho(\D)\limu{|\Delta|\to\infty}  & 2^\frac{2\delta+d+1}{2} \pi^\frac{d-3}{2} c \,\sin(\pi \delta) (-\D^2)^{\delta-\hd} \int_0^\infty d\nu \, \nu^{\frac{d+1}{2}-2\delta} K_{\frac{d-1}{2}}(\nu) \\
&=c\frac{\pi^\frac{d+1}{2}(-\D^2)^{\delta-\hd}}{2^{\delta-d-1} \Gamma(\delta)\Gamma(\delta-\hd+\half)}~.
\ea
and in particular for $\D \in \CP$ one finds:
\be\label{eq:largeD rho}
\tcbhighmath[boxrule=-1pt,arc=3pt,colback=gray!30!white]{
\rho(\D)\limu{\lambda\to\infty} \frac{  (4\pi)^\frac{d+1}{2}c}{2^{\delta}\Gamma(\delta)\Gamma(\delta-\hd+\half)}  |\lambda|^{2\delta-d}}
\ee
where we again use the notation $\D=\hd+i\lambda$.  This is, in fact, the large $\D$ limit of the spectral density of a CFT in the bulk as one expects when assuming a flow to a CFT in the UV. One can also use the large $\D$ limit of the free two-point function in~\reef{eq: large Delta lim} to find the large $\D$ limit of \Kallen decomposition~\reef{eq:wave function}. It is straightforward to check that~\reef{eq:largeD rho} matches with the large $\D$ limit of $\Braket{\phi^2\phi^2}$ spectral density found in~\cite{Bros:2009bz,Hogervorst:2021uvp,Loparco:2023rug}. 

Now one can simply use the large $\D$ limit of $\frak{g}_\D$ in~\reef{eq: g limit} to find
\be
a_\D \limu{\lambda\to\pm\infty}  \frac{   \pi^\frac{1}{4} e^{\pm\frac{i \pi d}{8}} 2^\frac{\delta+d+1}{2}}{\sqrt{\Gamma(\delta)\Gamma(\delta-\hd+\half)}}  |\lambda|^{\delta-\frac{d-2}{4}} e^{-\frac{\pi}{2} |\lambda|}~,
\ee
where the $\pm$ is representing the direction of the limit to infinity on principal series and $\delta$ is the scaling dimension of the bulk field in UV. As one can see, by normalizing the boundary operators to have a unit CFT two-point function (i.e. $\alpha_\D=1$), the large $\D$ limit of the bulk-to-boundary expansion is universally exponentially decaying. 

\newpage
\section{Inversion formula: boundary operators from the bulk}\label{sec:Inversion formula}
The AdS boundary-operator/bulk-state correspondence provides a procedure for constructing the boundary operators. In other words, given a bulk theory and insertion of a bulk field at some point, one can define a set of states on each global time slice. Then by evolving back in time (shrinking the hemisphere in figure~\ref{fig:AdS}), one reaches a set of local boundary operators. 

In this section, we provide such a construction for the late-time boundary of de Sitter spacetime.  Given a bulk theory with field $\phi$, we can define a continuous (in $\D$) set of boundary operators $\Oo_\D(\V y)$. In other words, we would like to invert the bulk-to-boundary expansion~\reef{eq:the expansion} where the bulk field is expanded in terms of boundary operators. Roughly speaking, to find such an inversion formula is to define a boundary operator $\Oo_\D (P)$ as an integration of the bulk field $\phi(Y)$  against a non-local kernel $\frak{K}(P,Y)$ over some region of the bulk:
\be\label{eq: basic inversion}
\Oo_\D(P)= \frac{1}{\ND{\D}} \int_\text{bulk} dY\, \frak{K}(Y,P) \phi (Y)~.
\ee
Before deriving this formula and finding the explicit expressions for $N_\D$ and the kernel $\frak{K}$, let us provide a heuristic argument about what we expect $\frak{K}$ to be. Since the operator $\Oo(P)$ has scaling dimension $\D$, $\frak{K}$  must  scale as $\Lambda^{-\D}$ upon rescaling of embedding space boundary point $P\to\Lambda P$. Moreover, $\frak{K}$ can only be a function of $P\cdot Y$ -- \SOd-invariant product made out of two points $Y$ and $P$. Therefore, one finds that $\frak{K}$ is proportional to the bulk-to-boundary propagator:
\be
 \frak{K}(Y,P) \propto\frac{1} {(-2P\cdot Y)^\D}=K_\D(Y,P)~.
\ee
The factor "$-2$" above is picked for later convenience\footnote{A positive number instead of $-2$ would flip the $i\e$ prescription.}.
\subsection{Direct derivation}\label{sec:proof}
We start with the bulk-to-boundary expansion~\reef{eq:the expansion} and integrate it against the bulk-to-boundary propagator:
\be
\int_{\eta,\V{y}} K^\pm_{\D}(\eta,\V{y};\V{y}') \phi(\eta,\V{y}) = \int^\infty_{-\infty} d\lambda' a_{\D'}\int_{\eta,\V{y}} K^\pm_{\D}(\eta,\V{y};\V{y}') \,  \hat{D}_{\D'}(\eta,\V{y}) \Oo_{\D'}(\V{y})~,
\ee
where we use the shorthand notation defined in~\reef{eq: dS volume int} for the integration over the de Sitter volume. Here, we keep track of the $ i\epsilon$ prescription for the causally connected regions using a superscript on the bulk-to-boundary propagator defined in~\reef{eq: K i eps}. 
Next, we perform an integral by parts for the $y$-derivatives in $ \hat{D}_{\D'}(\eta,\V{y})$\footnote{Alternatively, one can first perform a Fourier transform and then the integral by parts to avoid the subtlety of Taylor expansion for $|\eta|>|\V{y}|$.} and find
\be\label{eq: derivatrion 2}
\int_{\eta,\V{y}} K^\pm_{\D}(\eta,\V{y};\V{y}') \phi(\eta,\V{y}) =  \int^\infty_{-\infty} d\lambda'  a_{\D'} \int_{\V{y}}  \Oo_{\D'}(\V{y})\int_{\eta}  \hat{D}_{\D'}(\eta,\V{y})  K^\pm_{\D}(\eta,\V{y};\V{y}')~.
\ee
In what comes, we show that 
\be\label{eq: int of D and K}
\mathcal{I^\pm} = \int \frac{d\eta}{(-\eta)^{d+1}}  \hat{D}_{\D'}(\eta,\V{y})  K^\pm_{\D}(\eta,\V{y};\V{y}') = \bbc{}^\pm \delta_{\lambda,\lambda'} \delta_{\V{y},\V{y}'}~,
\ee
for some position-independent function $\bbc{}^\pm$ that will be determined shortly. 
Exploiting the explicit translation invariance in the bulk-to-boundary propagator, we can, without loss of generality, set $\V{y}' = 0$.
First, one can perform a Fourier transformation with respect to $\V{y}$:
\ba
\tilde{\mathcal{I}}^\pm&= \int_{\eta} \int_{\V{y}} e^{i\V{k}\cdot\V{y}} \hat{D}_{\D'}(\eta,\V{y})  K^\pm_{\D}(\eta,\V{y};0) \\
&= \int_{\eta} (-\eta)^{\D'} \int_{\V{y}} e^{i\V{k}\cdot\V{y}} \, _0F_1(\D'-\hd+1,-\frac{1}{4}\eta^2k^2)  K^\pm_{\D}(\eta,\V{y};0) \\
&= \int_{\eta} (-\eta)^{\hd} 2^{i\lambda'} k^{-i\lambda'} \Gamma(i\lambda'+1) J_{i\lambda'}(-k\eta)\int_{\V{y}}\,   e^{i\V{k}\cdot\V{y}}  K^\pm_{\D}(\eta,\V{y};0) \\
&= \mp e^{\mp\pi \lambda} \frac{\pi^{\hd+1}\lambda' \Gamma(i\lambda')}{2^{i(\lambda-\lambda')}\Gamma(\hd+i\lambda)} \,k^{i(\lambda-\lambda')}\int^0_{-\infty} \frac{d\eta}{-\eta}  H^{(1)/(2)}_{i\lambda}(-k\eta) J_{i\lambda'}(-k\eta)~,
\ea
where we performed an integral by parts to find the second line, applied the hypergeometric and Bessel function relation~\reef{eq:0F1 to J} to find the third line and finally used the Fourier transform of the bulk-to-boundary propagator~\reef{eq: FT of Kpm} to find the last line. Here by $ H^{(1)/(2)}$ we mean that $ H^{(1)}$ corresponds to $\tilde{\mathcal{I}}^+$  and $ H^{(2)}$ corresponds to $\tilde{\mathcal{I}}^-$.
The last integral is an alternative version of the Kontorovich–Lebedev transform~\cite{Jones1980TheKT}. The conventional form of the Kontorovich–Lebedev transform leads to an orthogonality relation of Bessel K functions  -- see~\cite[A.18]{Laddha:2022nmj}\cite{szmytkowski2009}.  The relevant version of the Kontorovich–Lebedev transform can be found in~\cite{KLtrans,jones1964electromagnetism} and is given by:
\be
F(\lambda)=\int_0^\infty dx\, H^{(1)/(2)}_{i\lambda}(x) f(x)~, \qquad f(x) =\frac{1}{2x} \int^{\infty}_{-\infty} d\lambda \, \lambda \,J_{i\lambda} (x) F(\lambda)~,
\ee
which leads to the desired orthogonality relation:
\be\label{eq: Kontorovich–Lebedev}
\int_0^\infty \frac{dx}{x} J_{i\lambda}(x) H^{(1)/(2)}_{i\lambda'}(x) =\mp\frac{2}{\lambda} \delta(\lambda-\lambda')~.
\ee
Using this relation, we find that $\tilde{\mathcal{I}}^\pm$ is just a constant (there is no $k$-dependence left)
and hence the the Fourier transform back is a delta function in position space. So we end up with~\reef{eq: int of D and K} in which
\be
\bbc{}^\pm =2\pi \xi_\D^{\pm}~.
\ee

Now that we derived~\reef{eq: int of D and K}, we can simply plug it back into~\reef{eq: derivatrion 2} to find our inversion formula:

\be\label{eq: the inversion formula}
\tcbhighmath[boxrule=-1pt,arc=3pt,colback=gray!30!white]{
\Oo_\D(\V{y})= \frac{1}{\ND{\D}} \int_{\eta',\V{y}'}K^\pm_{\D}(\eta',\V{y}';\V{y}) \phi (\eta',\V {y}')
}
\ee
where
\be\label{eq: N and xi}
N_\D =2\pi \xi_\D^{\pm} a_\D~.
\ee
Again, the $\pm$ stands for the choice of the $i\e$ prescription for the bulk-to-boundary propagator and it matches with the  corresponding $i\e$ prescription of the bulk field on in-in (Keldysh-Schwinger) contour. In other words, if the bulk time corresponds to a right(left) vertex, its conformal time acquires imaginary part $\eta\to\eta+(-) i\e$. Notice that the normalization is simply given by the spectral density through equation~\reef{eq: gamma def}. One important remark is that the integral above is over both timelike and  spacelike separation, getting non-zero contributions from causally disconnected region. It would be interesting to understand the physical interpretation of this. 

As a result of the discussion in Section~\ref{sec:AdS expansion}, we began with the bulk-to-boundary expansion and later derived the inversion formula. However, the reverse approach could also work: one could start with the inversion formula and derive the bulk-to-boundary expansion. This is essentially what we do in the next section.
\begin{figure}[t!]
\centering
\begin{tikzpicture}[line width=1. pt, scale=2]

\draw[lightgray, line width=1.pt] (-0.5,0) -- (0.5,0.75);
\draw[lightgray, line width=2.pt] (-1,0.75) -- (1,0.75);

\draw  (0.5,0.75) node[fill, circle, scale=0.5]  {};
\draw[fill=black] (-0.5,0) circle (.03cm);

\node[scale=1] at (0.5,0.9){$\Oo_\D(P)$};
\node[scale=1] at (0.1,0.25){$K_\D$};
\node[scale=1] at (-0.5,-.2) {$\displaystyle\int_Y \phi(Y)$};
\end{tikzpicture}
   \caption{The boundary operator $\Oo(P)$ with dimension $\D$ is given by an integral over the bulk field $\phi(Y)$ multiplied by the bulk-to-boundary propagator with the same scaling dimension. }
  \label{fig:inversion formula }
\end{figure}
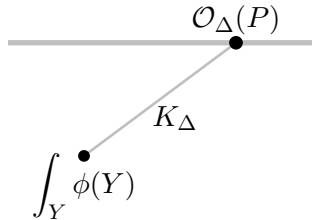
\subsection{Alternative derivation}\label{sec:Alternative}
In the previous section, we started with~\reef{eq:the expansion} and inverted it. In this section, we will see that our inversion formula satisfies the bulk-to-boundary expansion. In other words, we will start with~\reef{eq: the inversion formula} and derive the bulk-to-boundary expansion meaning that we could take~\reef{eq: the inversion formula} as the definition of boundary operators and derive~\reef{eq:the expansion}.

Consider~\reef{eq: the inversion formula} and act on both sides with the boundary-to-bulk connector, integrating against a kernel $a_\D$ over the principal series:
\ba
 \int_{-\infty}^\infty d\lambda~a_\D \Dh{}{}\Oo_\D(\V{y})= \int_{\eta',\V{y}'}  \phi (\eta',\V {y}') \int_{-\infty}^\infty d\lambda~  \frac{a_\D}{\ND{\D}} \Dh{}{}  K^\pm_{\D}(\eta',\V{y}';\V{y})~.
\ea
Although we now assume that $a_\D$ is a generic function, it is going to be the same as the kernel in the bulk-to-boundary expansion. 
As we have seen in~\reef{eq: psi from D on K }, the action of the boundary-to-bulk connector on the bulk-to-boundary propagator is simply $\Pi_\D(\sigma)$. We restore the $i \e$ prescription by requiring both conformal times acquiring the same imaginary part:  $\eta \to \eta+ i\e $ and $\eta' \to \eta'+ i\e $ to find
\be\label{eq: alternate der}
 \int_{-\infty}^\infty d\lambda~a_\D\Dh{}{}\Oo_\D(\V{y})= \int_{\eta',\V{y}'}  \phi (\eta',\V {y}') \int_{-\infty}^\infty d\lambda~  \frac{a_\D}{\ND{\D}}\Pi^\pm_\D(\sigma)~,
\ee
where $\Pi^\pm(\sigma)=\Pi(\sigma\pm i\e)$. In appendix~\ref{sec: Harmonic analysis}, we find that $\Pi_\D^\pm (\sigma)$ satisfies a completeness relation 
\be\label{eq: dS pi completeness}
\int_{-\infty}^\infty d\lambda~ \frac{1}{2\pi\xi^\pm}\Pi_{\D}(\sigma^\pm)= \delta(Y_1,Y_2)~.
\ee
This completeness relation implies that if one asks
\be
\frac{a_\D}{\ND{\D}}=\frac{1}{2\pi\xi^\pm_\D}~,
\ee
then the integral on the RHS of~\reef{eq: alternate der} becomes trivial and one finds the desired bulk-to-boundary expansion~\reef{eq:the expansion}. Note that the relation above is exactly what we found in~\reef{eq: N and xi}.
\subsection{Consistency checks}\label{sec:Consistency}
In this section, we explicitly verify that the inversion formula reproduces the expected boundary two-point functions, including the contact terms. Furthermore, we verify that the inversion formula reproduces the known boundary operator correlation functions from perturbation theory. 
\subsubsection{Boundary two-point function}
Let us insert~\reef{eq: the inversion formula} into the boundary two-point function. 
\be
\Braket{\Oo_{\D_1}(\V y_1)\Oo_{\D_2}(\V y_2)} = \frac{1}{\ND{\D_1}\ND{\D_2}}\int_{Y'_1,Y'_2} K^\pm_{\D_1}(\eta'_1,\V{y}'_1;\V{y}_1)K^\mp_{\D_2}(\eta'_2,\V{y}'_2;\V{y}_2) \Braket{\phi (\eta'_1,\V y'_1)\phi (\eta'_2,\V y'_2)}~.
\ee
The choice of $i\e$ prescription will become clear shortly. 
Now we use the \Kallen spectral decomposition~\reef{eq:Kallen} to rewrite the above expression in terms of free two-point function:
\ba\label{eq: OO and Kallen}
\Braket{\Oo_{\D_1}(\V y_1)\Oo_{\D_2}(\V y_2)} &=\int_\Real {d\lambda}  \frac{\rho(\D)}{\ND{\D_1}\ND{\D_2}} \int_{Y'_1,Y'_2} K^\pm_{\D_1}(\eta'_1,\V{y}'_1;\V{y}_1)K^\mp_{\D_2}(\eta'_2,\V{y}'_2;\V{y}_2) G_\D^{\text{free}}(\sigma')~,
\ea
where $\D=\hd+i\lambda$ and $\sigma'= Y'_1\cdot Y'_2$. Again, we do not spell out the contributions from other representation and keep them implicit. Note that the free Wightman two-point function comes with a definite $i\e$ prescription. In particular, the first and second conformal times should have the opposite $i\e$ prescription, see e.g. \cite{Loparco:2023rug,Sleight:2021plv} for more details.  

The final integral in~\reef{eq: OO and Kallen} is just a theory independent quantity fixed by symmetry. To calculate it,  we first use the split representation in de Sitter, given by~\reef{eq: ds split rep} and illustrated in figure~\ref{fig:Split representation}, to rewrite the free bulk two-point function in terms of two bulk-to-boundary propagators. The integral over spacetime in equation above then becomes:  
\ba
{e^{\pm\pi \lambda}} \frac{\Gamma(\D)\Gamma(d-\D)}{4 \pi^{d+1}}   \int_{\V y, Y'_1,Y'_2} K^\pm_{\D_1}(\eta'_1,\V{y}'_1;\V{y}_1) K^\pm_{\D}(\eta'_1,\V{y}'_1;\V{y})  K^\mp_{\D_2}(\eta'_2,\V{y}'_2;\V{y}_2) K^\mp_{\Db}(\eta'_2,\V{y}'_2;\V{y})~.
\ea 
Now, the integrals over $\int_{Y_1'}$ and $\int_{Y_2'}$ are given by the V diagram identity discussed in appendix~\ref{sec: V diagram}. In particular, using \reef{eq: KK V diagram} we find that the integral simplifies to:
\ba
\Braket{\Oo_{\D_1}(\V y_1)\Oo_{\D_2}(\V y_2)} &= \frac{2\pi^{\hd+2}\Gamma(i\lambda_1)}{\lambda_1\sinh(\pi \lambda_1) \Gamma(\hd+i\lambda_1)} \frac{\rho(\D_1)}{N_{\D_1}^\pm N_{\D_1}^\mp}\frac{\delta_{\lambda_1,\lambda_2}}{y_{12}^{2\D_1}} \\
&+ \frac{2\pi^{d+3}\delta_{\lambda_1,-\lambda_2} \delta_{\V y_1,\V y_2}}{\lambda_1^2 \sinh(\pi \lambda_1)^2 e^{\pm \pi \lambda_1} \Gamma(\hd+i\lambda_1)\Gamma(\hd-i\lambda_1)} \frac{\rho(\D_1)}{N_{\D_1}^\pm N_{\Db_1}^\mp}~.
\ea
Let us rewrite the equation above by expressing the inversion formula normalization in terms of the bulk-to-boundary coefficient, using equation~\reef{eq: N and xi}:
\ba\label{eq: Boundary two point from inversion}
\Braket{\Oo_{\D_1}(\V y_1)\Oo_{\D_2}(\V y_2)} &=  \alpha_{\D_1}\frac{\delta_{\lambda_1,\lambda_2}}{y_{12}^{2\D_1}}+\beta_{\D_1}\delta_{\lambda_1,-\lambda_2} \delta_{\V y_1,\V y_2}~,
\ea
where we introduce the boundary two-point functions normalizations:
\be\label{alpha and beta from inversion}
\alpha_\D=  \frac{\Gamma(\D) \Gamma(\hd-\D)}{2\pi^{\hd+1}}\frac{\rho(\D)}{a_{\D}^2}~, \qquad \beta_\D= \frac{e^{\pm\pi\lambda}\Gamma(\D-\hd)\Gamma(\hd-\D)}{2\pi} \frac{\rho(\D)}{a_{\D}a_{\Db}}~.
\ee
These match exactly with what we found by inserting the \Kallen decomposition into equations~\reef{eq: a and rho relation} and~\reef{eq: beta and rho relation}.
As before, one may also normalize the boundary operators by setting $\alpha_\D=1$, thereby fixing $a_\D$ in terms of the \Kallen spectral density to obtain~\reef{eq: gamma def}.
As seen explicitly from equation~\reef{eq: Boundary two point from inversion}, a byproduct of the inversion formula in eq.~\reef{eq: the inversion formula} is the existence of contact boundary terms, which naturally arise in this context.

\subsubsection{Recovering perturbation theory}\label{sec:Perturbation}
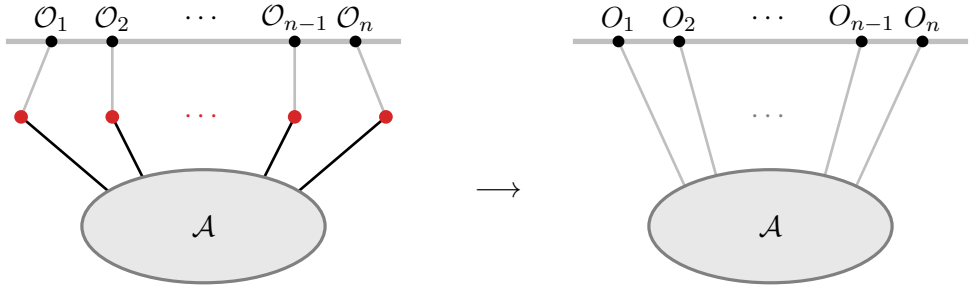
\begin{figure}[t!]
   \centering
           \begin{tabular}{cc}
\scalebox{1}{
\raisebox{-33pt}{
\begin{tikzpicture}[line width=1. pt, scale=2]
\draw[lightgray, line width=2.pt] (-1.3,0.6) -- (1.3,0.6);

\draw[lightgray, line width=1.pt]   (-1,0.6)   -- (-1.2,0.1);
\node[scale=1] at (-1,0.75) {$\Oo_{1}$};
\draw[fill=black]  (-1,0.6) circle (.03cm);
\draw[fill=black] (-1.2,0.1) -- (-0.5,-0.5);

\draw[lightgray, line width=1.pt]   (-0.6,0.6)   -- (-0.6,0.1);
\node[scale=1] at (-0.6,0.75) {$\Oo_{2}$};
\draw[fill=black]  (-0.6,0.6) circle (.03cm);
\draw[fill=black] (-0.6,0.1) -- (-0.3,-0.5);

\draw[lightgray, line width=1.pt]   (0.6,0.6)   -- (0.6,0.1);
\node[scale=1] at (0.6,0.75) {$\Oo_{n-1}$};
\draw[fill=black]  (0.6,0.6) circle (.03cm);
\draw[fill=black] (0.6,0.1) -- (0.3,-0.5);

\draw[lightgray, line width=1.pt]   (1,0.6)   -- (1.2,0.1);
\node[scale=1] at (1,0.75) {$\Oo_{n}$};
\draw[fill=black]  (1,0.6) circle (.03cm);
\draw[fill=black] (1.2,0.1) -- (0.5,-0.5);

\node[scale=1] at (0,0.75) {$\cdots$};
\node[scale=1,red3] at (0,0.1) {$\cdots$};

\draw  (-1.2,0.1)  node[red3,fill, circle, scale=0.5]  {};
\draw (-0.6,0.1) node[red3,fill, circle, scale=0.5]  {};
\draw  (1.2,0.1)  node[red3,fill, circle, scale=0.5]  {};
\draw (0.6,0.1) node[red3,fill, circle, scale=0.5]  {};

\filldraw[color=gray, fill={rgb:black,1;white,10}, very thick, scale=0.5](0,-1.25) ellipse[y radius=0.75cm, x radius=1.6cm] node[black]{ $\mathcal{A}$};
\end{tikzpicture}
}
}& ~~$\longrightarrow$~~
\scalebox{1}{
\raisebox{-33pt}{
\begin{tikzpicture}[line width=1. pt, scale=2]
\draw[lightgray, line width=2.pt] (-1.3,0.6) -- (1.3,0.6);

\draw[lightgray, line width=1.pt]   (-1,0.6)   -- (-0.5,-0.5);
\node[scale=1] at (-1,0.75) {$\O_{1}$};
\draw[fill=black]  (-1,0.6) circle (.03cm);

\draw[lightgray, line width=1.pt]   (-0.6,0.6)   -- (-0.3,-0.5);
\node[scale=1] at (-0.6,0.75) {$\O_{2}$};
\draw[fill=black]  (-0.6,0.6) circle (.03cm);

\draw[lightgray, line width=1.pt]   (0.6,0.6)   -- (0.3,-0.5);
\node[scale=1] at (0.6,0.75) {$\O_{n-1}$};
\draw[fill=black]  (0.6,0.6) circle (.03cm);

\draw[lightgray, line width=1.pt]   (1,0.6)   -- (0.5,-0.5);
\node[scale=1] at (1,0.75) {$\O_{n}$};
\draw[fill=black]  (1,0.6) circle (.03cm);

\node[scale=1] at (0,0.75) {$\cdots$};
\node[scale=1,gray] at (0,0.1) {$\cdots$};
\filldraw[color=gray, fill={rgb:black,1;white,10}, very thick, scale=0.5](0,-1.25) ellipse[y radius=0.75cm, x radius=1.6cm] node[black]{ $\mathcal{A}$};
\end{tikzpicture}
}
}
\end{tabular}
   \caption{Recovering perturbation theory using the inversion formula. Left: Each boundary operator $\Oo_i$ is given by integrating over the bulk point (red) multiplied by the bulk-to-boundary propagator (gray). Using the broken leg identity in figure~\ref{fig:broken leg diagram}, the integral simplifies to a bulk-to-boundary propagator, thereby recovering the perturbation theory in the right figure corresponding to equation~\reef{eq: discrete from bulk}.}
  \label{fig:PT}
\end{figure}

There has been growing interest in perturbative calculations for de Sitter in-in correlators over the last decade, either directly or through calculating the wave function coefficients, see, for example~\cite{Arkani-Hamed:2015bza,Arkani-Hamed:2018kmz,Goodhew:2020hob,Sleight:2021plv,Arkani-Hamed:2023kig}. The boundary correlators in these perturbative calculations are made of operators which are derived from the boundary expansion in~\reef{eq: free expansion}. In other words, a boundary correlator can be derived by pushing the bulk correlator to the boundary and substituting the external lines with the bulk-to-boundary propagator with the appropriate normalization:
\be\label{eq: discrete from bulk}
\Braket{\O_1(\V y_1) \cdots \O_n(\V y_n)} \sim \lim_{\eta_i\to0^-} \Braket{\phi_1(\eta_1,\V y_1) \cdots \phi_n(\eta_n,\V y_n)}~\\
\sim \prod_{i=1}^n K_{\D_i}(Y''_i,\V y_i) \mathcal{A}(Y''_i)~,
\ee
where we refer to  the contribution from the bulk interaction indicated by the gray box in figure~\ref{fig:PT} as $\mathcal{A}(Y''_i)$ and  the use  of $\sim$ means that we left the normalizations implicit.
In what comes, we will show that the correlation function of the boundary operators defined by the inversion formula~\reef{eq: the inversion formula}, indeed reproduces the above correlator. Before doing so, let us make a comment about how one should adapt the language of the continuous boundary operator in~\reef{eq:the expansion} and \reef{eq: the inversion formula}, denoted by the curly letter $\Oo_{\D}$, to the case of the discrete free boundary operator $O_{\D^*}$ in~\reef{eq: free expansion} where $\D^*(d-\D^*)=m^2$. One can interpret the operators $O_{\D^*}$ as a very narrow distribution of $\Oo_{\D}$ around $\D^*$:
\be\label{eq:zakhar}
\int_{\lambda^*-\e}^{\lambda^*+\e}  d\lambda \; a_\D\; (-\eta)^{\D}\Oo_{\D}(\V y) = (-\eta)^{\D^*}\sqrt{\frak{g}_{\D^*}} O_{\D^*}~.
\ee
where we defined $\D^* =\hd+i \lambda^*$ with $\lambda^*\geq0$. So in practice, one can simply substitute $a_\D$ with $ \sqrt{\frak{g}_{\D^*}} \delta_{\lambda,\lambda^*}$ when going from the continuous operators to discrete operators. Same is true for the shadow operator in~\reef{eq: free expansion}.

Now, let us go back to the calculation of a boundary correlator $\Braket{\Oo_1(\V y_1) \cdots \Oo_n(\V y_n)}$  corresponding to a set of  bulk fields $\{\phi_i\}$ with dimension $\D^*_i$ at position $Y_i$ with a bulk correlator that is calculated in advance, say through the in-in formalism  $\Braket{\phi_1(\eta_1,\V y_1) \cdots \phi_n(\eta_n,\V y_n)}_{\text{in-in}}$. Using the inversion formula one finds:
\be
\Braket{\Oo_1(\V y_1) \cdots \Oo_n(\V y_n)} = \frac{1}{\prod_{i=1}^n (2\pi\xi_{\D_i} a_{\D_i})}\int_{Y'_i}\Braket{\phi_1(Y'_1) \cdots \phi_n(Y'_n)}_{\text{in-in}} \prod_{i=1}^n K_{\D_i}(Y'_i;\V y_i)~,
\ee
where $Y_i' =(\eta'_i,\V y'_i)$ is embedding space notation for the bulk points and $\D_i$ is the dimension of boundary operator $\Oo_i$. As illustrated in figure~\ref{fig:PT}, the in-in correlator $\Braket{\phi_1(Y'_1) \cdots \phi_n(Y'_n)}$ is calculated through some arbitrary bulk interaction, and each $\phi_i(Y'_i)$ has an extra line attached to it. These lines, indicated by solid black lines in figure~\ref{fig:PT}, are the bulk-to-bulk propagator $G_{\D^*_i}^\text{free}(\sigma)$\footnote{Here we left all the $i\e$ prescriptions implicit. 
More precisely, the in-in correlator $\Braket{O_1(\V y_1) \cdots O_n(\V y_n)} $  has a sum over the right and left vertices~\cite{Weinberg:2005vy,DiPietro:2021sjt,Sleight:2021plv}, and the bulk-to-boundary propagators (the gray lines in figure~\ref{fig:PT}) have the same $i\e$ prescription as the interaction vertex they are attaching to~\cite{DiPietro:2021sjt} -- a right vertex corresponds to $K^+_\D$ and a left vertex corresponds to $K^-_\D$.  By taking the solid black lines in figure~\ref{fig:PT} to be the Wightman function $G^\text{free}_\D$ (neither time-ordered nor anti time-ordered), we fix the $i\e$ prescription of $\phi_i(Y'_i)$ to be the opposite of the interaction vertex they are connecting to. When using the broken leg identity, the $i\e$ prescription flips again and matches with the interaction vertex, consistent with the  in-in  calculation of $\Braket{O_1(\V y_1) \cdots O_n(\V y_n)} $.}.
The combination  of each  $G_{\D^*_i}^\text{free}(\sigma)$ and bulk-to-boundary propagator produces the broken leg integral found in~\reef{eq:broken leg id} which simplifies the integral as follows:
\ba\label{eq: broken leg continuous discrete}
\Braket{\Oo_{1}(\V y_1) \cdots \Oo_{n}(\V y_n)}  &= \frac{1}{\prod_{i=1}^n (2\pi\xi_{\D_i} a_{\D_i})}\int_{Y'_i} \prod_{i=1}^n G_{\D^*_i}(Y''_i,Y'_i) K_{\D_i}(Y'_i;\V y_i) \times \mathcal{A}(Y''_i)\\
&= \prod_{i=1}^n \frac{\Gamma(\hd+i\lambda_i)\Gamma(-i\lambda_i)}{4\pi^{\hd+1} a_{\D_i}}\left( \delta_{\lambda_i,\lambda_i^*}+ \delta_{\lambda_i,-\lambda_i^*}\right)  K_{\D_i}(Y''_i;\V y_i) \times\mathcal{A}(Y''_i)~.
\ea 
Now by using the dictionary between discrete and continuous operators discussed above, it is straightforward to see that the correlator above produces the perturbative discrete correlator defined in~\reef{eq: discrete from bulk}:
\be
\Braket{\Oo_1(\V y_1) \cdots \Oo_n(\V y_n)}  \to \Braket{O_1(\V y_1) \cdots O_n(\V y_n)}~.
\ee
Note that, depending on whether one wants to calculate the correlator of $\O_i$ or $\tilde{\O}_i$, one of the delta functions in~\reef{eq: broken leg continuous discrete} becomes relevant.

\subsection*{A three-point function example}
Three-point functions in a conformal theory are fixed by the conformal Ward identities up to a normalization i.e.~the OPE coefficient. A boundary three-point function made out of our inversion formula, by construction, would satisfy the conformal Ward identities and produce the conformal three-point structure. Here, we illustrate this  fact by an explicit computation for a simple example.
More concretely, we consider a scalar free massive theory with two operators $\phi_1$ and $\phi_2$ with dimensions $\D_{\phi_1}=\hd+i\lambda_1$ and $\D_{\phi_2}=\hd+i\lambda_2$ on principal series with $\lambda_i>0$. We focus on a boundary three-point function $\Oo_{\phi^2}$ that corresponds to the composite bulk operator $\phi_1 \phi_2 (Y)$ and two boundary fields $\Oo_\phi$ that correspond to bulk operator $\phi(Y)$. The three point-function according to the inversion formula is given by
\be\label{eq: start 3point}
\Braket{\Oo_{\phi}(\V y_1)\Oo_{\phi}(\V y_2){\Oo}_{\phi^2}(\V y_3)} = \frac{1}{\prod_i N_{\D_i}}  \int_{Y'_i}  \Braket{\phi_1(Y_1')\phi_2(Y_2')\phi_1\phi_2(Y_3')} \prod_i  K_{\D_i}(Y_i',\V y_i)~,
\ee
where the products are over the corresponding field on LHS. We also left the $i\e$ prescription of the bulk-to-boundary propagators implicit but they can easily be recovered considering the explicit expression of the bulk three-point function we spell out next. The bulk three-point function is simply given by the product of two bulk-to-bulk two-point functions:
\be
 \Braket{\phi_1(Y_1')\phi_2(Y_2')\phi_1\phi_2(Y_3')} =G_{\D_{\phi_1}}(Y'_1,Y'_3)G_{\D_{\phi_2}}(Y'_2,Y'_3)~.
\ee
Upon substituting this back into~\reef{eq: start 3point} and using the broken leg identity~\reef{eq:broken leg id}, one finds that the RHS of~\reef{eq: start 3point} is :
\ba
\Braket{\Oo_{\phi}(\V y_1)\Oo_{\phi}(\V y_2){\Oo}_{\phi^2}(\V y_3)}  & \propto \int_Y K_{\D_{\phi_1}}(Y,\V y_1) K_{\D_{\phi_2}}(Y,\V y_2)K_{\D_{\phi^2}}(Y,\V y_3)~,
\ea
where $\D_{\phi^2}$ is the dimension of the boundary operator ${\Oo}_{\phi^2}(\V y_1)$ and we did not spell out the coefficient to avoid clutter but it can be easily recovered using equation~\reef{eq: N and xi}.  The integral above is nothing but the three-point contact diagram~\cite{Freedman:1998tz,Sleight:2019mgd} which leads to the following known CFT three-point structure:
\be
\Braket{\Oo_{\phi}(\V y_1)\Oo_{\phi}(\V y_2){\Oo}_{\phi^2}(\V y_3)}  \propto \frac{1}{y_{12}^{\D_{\phi_1}+\D_{\phi_2}-\D}y_{13}^{\D_{\phi_1}+\D-\D_{\phi_2}}y_{23}^{\D_{\phi_2}+\D-\D_{\phi_1}}}~.
\ee
\subsection{Shadow transformation and hermitian conjugation}\label{sec:dagger}
A shadow transformation in a standard CFT produces an operator with a scaling dimension $\Db=d-\D$. In what follows, we will show that by taking a shadow transformation of our inversion formula, we land on the same boundary operator with shadow dimension. To see this, let us perform a shadow transformation defined as follows:
\be
S[\Oo_\D(\V y)] = \frac{1}{\mathcal{N}_\D} \int_{\V y'} \frac{1}{|\V y- \V y'|^{2\Db}}  \Oo_\D(\V y)~,
\ee
on the boundary operator for some normalization factor $\mathcal{N}_\D$. It is straightforward to verify that $S[\Oo_\D(\V y)] $ has scaling dimension $\Db$. We normalize the shadow transformation such that two shadow transformations yield the same operator $S^2=\mathbb{1}$. Using the bubble diagram in section~\ref{sec:bubblediagram}, this condition translates to the following relation:
\be\label{eq: NN B}
\mathcal{N}_\D\mathcal{N}_{\Db} = B_\D~.
\ee
Let us now perform the shadow transform on the inversion formula:
\be
S[\Oo_\D(\V y)]  =  \frac{1}{\mathcal{N}_\D \ND{\D}} \int_{\eta',\V y'}\phi (\eta',\V {y}') \int_{\V y''} \frac{K^\pm_{\D}(\eta',\V{y}';\V{y}'') }{|\V y- \V y''|^{2\Db}}~.
\ee
The last integral can be easily computed using either the late-time limit of split representation~\reef{eq: ds split rep} or alternatively rewriting the bulk-to-boundary propagator using the boundary-to-bulk connector:
\be
\int_{\V y} \frac{K^\pm_{\D}(\eta_2,\V{y}_2;\V{y}) }{|\V y- \V y_1|^{2\Db}}= \xi^{\pm}_{\D}K^\pm_{\Db}(\eta_2,\V{y}_2;\V{y}_1)~.
\ee
Using this identity, one finds the shadow transformation of $\Oo_\D$ is $\Oo_{\Db}$, up to a normalization:
\be
S[\Oo_\D(\V y)] =  \frac{ \xi^{\pm}_{\D}}{\mathcal{N}_\D \ND{\D}} \int_{\eta',\V y'}\phi (\eta',\V {y}') K^\pm_{\Db}(\eta',\V{y}';\V{y})
= \frac{ \xi^{\pm}_{\Db} \sqrt{\frak{g}_{\Db}}}{\mathcal{N}_\D \sqrt{\frak{g}_{\D}}}  \Oo_{\Db}(\V y) ~.
\ee
where we used the explicit expression for $N_\D$, and $a_\D$ from equations~\reef{eq: N and xi} and \reef{eq: gamma def}. We may now pick the normalization of the shadow transform such that the above factor becomes the identity. In other words by picking 
\be
\mathcal{N}_\D = \xi^{\pm}_{\Db}  \sqrt{\frac{\frak{g}_{\Db}} {\frak{g}_{\D}}}~,
\ee
the shadow transform is just switching the boundary dimension to its shadow:
\be
S[\Oo_\D(\V y)]  = \Oo_{\Db}(\V y)~.
\ee
Note that the normalization above satisfies~\reef{eq: NN B}.

The notion of hermitian conjugation of an operator is inherently based on its action on the states of the Hilbert space. 
The inversion formula~\reef{eq: the inversion formula}, makes such action manifest in terms of the bulk operators.
In particular, for a bulk theory with hermitian operators, $\phi^\dagger = \phi$, the inversion formula gives:
\be
\Oo_\D(\V{y})= \frac{1}{\ND{\D}} \int_{Y'}K^\pm_{\D}(Y';\V{y}) \phi (Y') \quad \longrightarrow \quad \Oo^\dagger_\D(\V{y})= \frac{1}{\ND{\D}^*} \int_{Y'}K^\mp_{\Db}(Y';\V{y}) \phi (Y')~.
\ee
This means that the hermitian conjugate of the boundary operator, up to a normalization, is the same as the boundary operator with the shadow dimension. However, there exists a subtlety  involving the $i \e$ prescription in equation above. As is apparent from the explicit expression of the bulk-to-boundary propagator, complex conjugation flips its $i\e$ prescription and so does the  hermitian conjugate in the inversion formula. So far, we have not discussed which $i \e$ prescription should be chosen for the boundary operators, however, the hermicity of the bulk-to-boundary expansion~\reef{eq:the expansion}, requires the existence of both $i\e$ prescriptions. A possible scenario is to take  a linear combination of both $i\e$ prescriptions as the boundary operators:
\be
\Oo_\D(\V y) = \zeta_+ \Oo_\D^+(\V y) +  \zeta_- \Oo_\D^-(\V y)~,
\ee
with 
\be\label{eq: O+ O_}
 \Oo_\D^+(\V y) =  \frac{1}{2\pi \xi_\D^+ a_\D} \int_{Y'}K^+_{\D}(Y';\V{y}) \phi (Y')~,\qquad \Oo_\D^-(\V y) =  \frac{1}{2\pi \xi_\D^- a_\D} \int_{Y'}K^-_{\D}(Y';\V{y}) \phi (Y')~.
\ee
Equation~\reef{eq: alternate der} requires that the sum of these coefficients equals unity: $\zeta_++\zeta_-=1$.
It is worth mentioning that in equation~\reef{eq: OO and Kallen}, we calculated $\Braket{\Oo^\pm\Oo^\mp}$. To compute $\Braket{\Oo^\pm\Oo^\pm}$, one needs to use the (anti-)time ordered free two-point function in~\reef{eq: OO and Kallen}.

It is also easy to see from~\reef{eq: O+ O_} that this set of boundary operators are related by hermitian conjugation:
\be
\left(\Oo_\D^\pm\right)^\dagger = \frac{N^\mp_{\Db}}{\left(N_\D^\pm\right)^*} \Oo^\mp_{\Db}= \Oo^\mp_{\Db}~.
\ee
The coefficients $\zeta^\pm$ are also constrained by hermicity of $\phi$. To see this, use the equation above and the bulk-to-boundary expansion to find the bulk operator hermitian conjugate:
\be
\phi^\dagger (\eta,\V y)= \int^{\infty}_{-\infty} d\lambda\; \zeta_+^*\frac{N^-_{\Db}}{\left(N_\D^+\right)^*}\; a_\D^* \Dhb{}{} \Oo^-_{\Db} (\V y) \quad + \quad (+\longleftrightarrow -)~,
\ee
where by $(+\longleftrightarrow -)$ we mean the swapping of the pluses and minuses in superscript and subscripts. Now one can rewrite the integral above by sending $\lambda\to-\lambda$ to find:
\be
 \phi^\dagger (\eta,\V y)=\int^{\infty}_{-\infty} d\lambda\; \zeta_-^* a_\D \Dh{}{} \Oo^+_{\D} (\V y) \quad + \quad (+\longleftrightarrow -)
\ee
where we use the explicit expression of inversion formula normalization in~\reef{eq: N and xi}. Then the hermicity of bulk field immediately implies that
\be
\zeta_\pm^*= \zeta_\mp~.
\ee
which considering the relation  $\zeta_++\zeta_-=1$, fixes the real part of these coefficients to be $\half$.  

In summary, strictly speaking, the hermiticity of the bulk theory requires the bulk-to-boundary expansion to include a linear combination of both $i\e$ prescriptions, with coefficients that are complex conjugates of each other and have a fixed real part.
It would be interesting to understand the full constraints on the coefficients of this linear combination. Perhaps, considering the global patch could be useful in this regard.
\newpage
\section{Conclusion and future directions}\label{sec:discusion}
The late-time boundary of de Sitter is conformal.  A feature that has been exploited extensively to study the perturbative inflationary correlators in de Sitter. 
However, a non-perturbative understanding of the QFT in de Sitter is still in early stages.  An approach that has been proven to be very useful for example when dealing with the IR divergences~\cite{Polyakov:2012uc,Gorbenko:2019rza,Anninos:2024fty} or understanding the necessity of the  boundary contact  terms from perspective of unitarity~\cite{Hogervorst:2021uvp}. Add to this the potential applications of the conformal bootstrap to put bounds on the cosmological observables.

In this work, we propose a non-perturbative definition of the boundary operators based on two main ingredients: the Hilbert space decomposition in irreducible unitary representations and the continuous spectrum of QFT in de Sitter.
At first, we discuss the complications and subtleties of a naive AdS-like approach to de Sitter boundary and then treat de Sitter as it is, a theory with a continuous spectrum.
We introduce a boundary theory made of a continuous family of  primary operators whose correlation functions satisfy the  Ward identities. These boundary operators are the outcome of pushing the bulk fields to the boundary -- the bulk-to-boundary expansion presented in~\reef{eq:the expansion}. 

We also found an inversion formula~\reef{eq: the inversion formula} to the bulk-to-boundary expansion that gives a recipe to construct the boundary operators from bulk operators by an integration against the bulk-to-boundary propagator. 
We examine this formula by recovering the expected boundary two-point function. A boundary two-point function that includes the standard CFT two-point function as well as the contact terms. We also use the inversion formula to recover the perturbation theory as a limit of bulk fields being pushed to the boundary.

There are many open questions and possible applications that are left for the future. Let us list some of them:

\begin{itemize}
\item In this work, we focused on the principal series contributions to the bulk-to-boundary expansion~\reef{eq:the expansion}; however, in general, there are contributions from other representations. It is interesting to find an inversion formula analog to~\reef{eq: the inversion formula} for other representations. We expect that complementary series contributions would be recovered by an analytic continuation in scaling dimension. On the other hand, from known examples of \Kallen spectral densities~\cite{Loparco:2023rug} and tensor product of one particles states~\cite{Penedones:2023uqc,Repka:1978}, we suspect complementary series show up as a discrete sum. Hence, a more detailed analysis is required. 
\item In this paper, we study the scalar fields. It would be interesting to generalize the results to spinning fields, conserved currents and gauge fields in this context. For instance, deriving an analog of the inversion formula for bulk photon and graviton would be interesting but requires  understanding the generalization of this construction to the other representations, the exceptional series. 

\item In section~\ref{sec:Convergence}, we derived the large $\D$ limit of the bulk-to-boundary coefficient and discussed its role in convergence of the bulk-to-boundary expansion inserted in correlation functions. However, we still lack a full understanding of such convergence as we do not know the large $\D$ limit of boundary OPE coefficients. The inversion formula could be useful to study such boundary OPE coefficients. For instance, in the case of a CFT in bulk, it is interesting to study how the boundary OPE coefficients are related to the bulk OPE coefficients.   
\item Our construction relies on the coordinate system we choose: the planar coordinates. On the other hand, it seems plausible to generalize the inversion formula to the global patch. It would be interesting to understand if a similar boundary operator construction like in~\reef{eq: the inversion formula} exists in the global patch. We expect the discussion of $i \e$ prescription in section~\ref{sec:dagger} would be related to an antipodal map $Y\to -Y$ on the global patch.
\item Most of the arguments in this paper can be formally generalized to AdS.
There might be interesting applications of an analog of the bulk-to-boundary expansion and inversion formula to AdS and flat-space holography.
\item It would be interesting to understand the analytic structure of the bulk-to-boundary coefficients. For the known examples like a CFT in bulk or free composite operators, the bulk-to-boundary coefficients have cuts in $\D$-plane due to the square root in~\reef{eq: gamma def}. It would be interesting to understand the physical interpretation of such cuts. 
\item The four-point functions of the boundary operators admit a partial wave expansion where unitarity implies positivity constraints on the partial wave coefficients~\cite{Hogervorst:2021uvp,DiPietro:2021sjt}. These four-point functions satisfy the crossing equation, resulting in bootstrap equations closely related to hyperbolic manifold bootstrap equations~\cite{Kravchuk:2021akc,Bonifacio:2023ban}. It would be interesting to adapt their method, especially taking into account the non-commutative feature of the boundary operators. 
\item Lastly, it is interesting to understand what we can learn from this construction for the case of dynamical gravity. In AdS, such generalization happens by the presence of the stress tensor at the boundary. Hence, studying the massless spin 2 field in de Sitter would be the first step towards this goal and might lead to hints for overcoming some of the difficulties that has been observed in this context~\cite{Strominger:2001pn,Maldacena:2002vr}.  
\end{itemize}


\section*{Acknowledgement}

I would like to thank Dio Anninos, Tarek Anous, Daniel Baumann, Miguel Correia, Jordan Cotler, Gabriel Cuomo, Victor Gorbenko, Aditya Hebbar, Shota Komatsu, Petr Kravchuk, Manuel Loparco, Dalimil Maz\'ač, Scott Melville, Prahar Mitra, Enrico Pajer, Jo\~ao Penedones, Suvrat Raju, Balt van Rees, Facundo Rost, Charlotte Sleight and Zimo Sun for useful suggestions and stimulating discussions.  I am also grateful to Tarek Anous, Aditya Hebbar,  Manuel Loparco, and Jo\~ao Penedones for their comments on the draft.
This work is supported by the European Research Council under the European Unions Seventh Framework Programme (FP7/2007-2013), ERC Grant agreement ADG 834878.
It is also supported by the European Union funding (ERC,  \raisebox{-2pt}{\includegraphics[height=0.9\baselineskip]{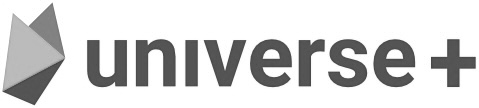}}, 101118787).
\newpage
\appendix
\section{Fourier transforms}\label{sec: Fourier}
In this appendix, we review the Fourier transform of several scalar functions. We denote the $d$-dimensional Fourier transform of $f(y)$ by $\tilde{f}(k)$ where we use $k=|\V k|$ denotes the norm of the vector. 
We define the Fourier transformation from position space to momentum space and back as follows:
\ba\label{eq: FT notation}
\widetilde{ \texttt{FT}} [\tilde {f}(k)]= f(y) =\int \frac{d^{d}k}{(2\pi)^{d}} e^{i\V{k}\cdot\V{y}}\tilde{f}(k)~, \qquad  {\texttt{FT}} [f(y)]=\tilde{f}(k)= \int d^{d}y \,\,e^{-i\V{k}\cdot\V{y}} f(y)~. \\
\ea

The  $d$-dimensional Fourier transform is related to the so-called Hankel Transform given by:
\be\label{eq: HT notation}
\texttt{HT}[f(z),z,x,\alpha]=\int^\infty_0 dz\,\, z\, J_\alpha (z x) \,f(z)~.
\ee
To see this relation, one simply needs to perform the angular part of the Fourier transform:
\ba\label{eq:FourierEuclidean}
 {\texttt{FT}} [f(y)]&= \int {d^{d}y}\,e^{-i\V{k}\cdot\V{y}} {f}(y)\\
&={\Omega_{d-2}}  \int y^{d-1}  dy\,  {f}(y)\int_{-1}^1 d\cos\theta \,(1-\cos^2\theta)^\frac{d-3}{2}  e^{-i k y \cos\theta}\\
&=\frac{(2\pi)^\hd}{k^{d-1}}\int_0^\infty dy\;  {f}(y) (k y)^{\frac{d}{2}} J_{\frac{d-2}{2}}(ky)\\
&= \frac{(2\pi)^{\frac{d}{2}}}{k^\frac{d-2}{2}} \texttt{HT[$ y^\frac{d-2}{2} {f}(y),y,k,{(d-2)}/{2}]$}~,
\ea
where $ \theta$ is the angle between the vectors $\V k$ and $\V y$, $\Omega_{d}=\frac{2\pi^\frac{d+1}{2}}{\Gamma(\frac{d+1}{2})}$ is the volume of $S^{d}$ and we used the Bessel function integral representation to go from the second to the third line. Now, one can use the many existing Hankel transformation tables as well as a built-in \texttt{Mathematica} function to perform the $d$-dimensional Fourier transform of a scalar function. 
\subsection{Action of bulk-to-boundary generator}\label{sec: D Fourier}
In this section, we verify equation~\reef{eq:DtoBulkBoundary} and~\reef{eq: psi from D on K }.
The method is to match the Fourier transform of the  both sides.
Without loss of generality we put $\V y_2$ at the origin and drop the index in $\eta_1$ and $\V y_1$. 
The Fourier transform of the first term in RHS of~\reef{eq:DtoBulkBoundary} is given by
\ba
\bold{D}_\D \equiv\Dh{}{}\frac{1}{|\V{y}|^{2\D}}~\longrightarrow~ 
\tilde{\bold{D}}_\D(k) &= 
(-\eta)^\D \int d^{d}y ~e^{-i\V{k}.\V{y}}  \,_0F_1(\D-\hd+1,\frac{1}{4}\eta^2\partial_{y}^2) \frac{1}{|\V{y}|^{2\D}}\\
&= (-\eta)^\D \,_0F_1(\D-\hd+1,-\frac{1}{4}\eta^2k^2) \int d^{d}y \frac{e^{-i\V{k}.\V{y}}}{|\V{y}|^{2\D}} \\
&= \frac{i\pi^\hd \lambda \Gamma( i \lambda) \Gamma(-i \lambda)}{2^{i\lambda}\Gamma(\hd+i\lambda)} (-\eta)^\hd  k^{i\lambda}J_{i\lambda}(-k\eta)~.
\ea
where we performed an integration by parts after expanding $_0F_1$ in the second\footnote{Strictly speaking, we are not really expanding. In fact we define the operation through this Fourier transformation.} line and used the identity:
\be\label{eq FT of power}
\int d^dy \frac{e^{i\V k \cdot \V y}}{y^{2a}} = \pi^\hd \frac{\Gamma(\hd-a)}{2^{2a-d}\Gamma(a)} k^{2a-d}~.
\ee
Similarly, one can find the second term on the RHS of~\reef{eq:DtoBulkBoundary}\footnote{We also used the fact that the hypergeometric $_0F_1$ can be written as the Bessel function:
\be\label{eq:0F1 to J}
_0F_1(a,-z)= \Gamma(a) z^\frac{1-a}{2} J_{a-1}(2\sqrt{z})~.
\ee}
\ba
\frak{D}_\D \equiv\Dh{}{}\delta^{(d)}(\V{y})~\longrightarrow~ 
\tilde{\frak{D}}_\D(k)&=(-\eta)^\D \int d^{d}y ~e^{-i\V{k}.\V{y}}  \,_0F_1(\D-\hd+1,\frac{1}{4}\eta^2\partial_{y}^2)\delta^{(d)}(\V{y})\\
&= (-\eta)^\D \,_0F_1(\D-\hd+1,-\frac{1}{4}\eta^2k^2)\\
&= 2^{i\lambda}\Gamma(i\lambda+1)(-\eta)^\hd k^{-i\lambda} J_{i\lambda}(-k\eta)~.
\ea
where, again, we performed an integral by parts and rewrote $_0F_1$ in terms of the Bessel function. The bulk-to-boundary propagator in momentum space is also given by:
\ba\label{eq: FT of Kpm}
\FT{K^{\pm}_\D}&=\pm i \frac{\pi^{\hd+1} (-\eta)^\hd k^{i\lambda}}{2^{i\lambda} e^{\pm \pi \lambda} \Gamma(\hd+i\lambda)} H^{(1)/(2)}_{i\lambda}(-k\eta)~.
\ea
To see this, one can first perform the Fourier transform of EAdS bulk-to-boundary $\frac{z^\D}{(z^2+\V x^2)^\D}$ which is proportional to modified Bessel function $K_\D$ and then analytically continue it to de Sitter using~\reef{eq: K to H}-- see~\cite{Loparco:2023rug} for more details.
As expected, \reef{eq: FT of Kpm} is nothing (up to normalization) but the familiar wave-function or mode-function in momentum space. 

Now, using the relation between Hankel functions and Bessel functions
\be
H_\nu^{(1)/(2)}(z)= \frac{J_{-\nu}(z) -e^{\mp i\pi\nu} J_\nu(z)}{\pm i \sin ( \pi\nu)}~,
\ee
it is straightforward to verify 
\be
 \FT{K^{\pm}_{\D}} = \tilde{\bold{D}}_\D(k) + \xi^{\pm}_{\D}\,\tilde{\frak{D}}_{\Db}(k)~,
\ee
where
\be
\xi^{\pm}_{\D}= \frac{\pi^\hd  \Gamma(i\lambda)}{\Gamma(\hd+i\lambda)e^{\pm \pi \lambda}}~.
\ee
Taking an inverse Fourier transform, one finds the desired relation~\reef{eq:DtoBulkBoundary}.

We now turn to~\reef{eq: psi from D on K }. Again, we use the translation invariance and set $\V{y}_1$ at the origin and call $y_{12}=y$ without loss of generality. The Fourier transform of the RHS after performing integration by parts is:
\ba
\text{FT}[\hat{D}_\D(\eta_2,\V{y}) K^{\pm}_\D(\eta_1,0,\V{y})]&= (-\eta_2)^\D \,{}_0F_1(\D-\hd+1,-\frac{1}{4}\eta_2^2k^2) \, \FT{K^{\pm}_\D}~,
\ea
where $\FT{K^{\pm}_\D}$ is given by~\reef{eq: FT of Kpm}.
We are now ready to perform the Fourier transform back to the positions space. 
Using~\reef{eq:0F1 to J} and \reef{eq:FourierEuclidean} we obtain the following Hankel transform:
\be\label{eq: K to H}
\mp\frac{\pi}{2^\hd}   \frac{ \lambda \Gamma(i\lambda)}{e^{\pm\pi \lambda}\Gamma(\hd+i\lambda)} \frac{(\eta_1 \eta_2)^\hd}{y^{\hd-1}} \texttt{HT}[k^\frac{d-2}{2}J_{i\lambda}(-k\eta_2)H^{(1)/(2)}_{i\lambda}(-k\eta_1),k,y,(d-2)/2]~.
\ee
Noting the relation between Hankel and modified Bessel functions:
\be
K_{\alpha }(x)=\begin{cases}
{\frac {\pi }{2}}i^{\alpha +1}H_{\alpha }^{(1)}(ix)&-\pi <\arg x\leq {\frac {\pi }{2}}\\{\frac {\pi }{2}}(-i)^{\alpha +1}H_{\alpha }^{(2)}(-ix)&-{\frac {\pi }{2}}<\arg x\leq \pi~,
\end{cases}
\ee
the Hankel transform above takes the form of the following integral identity~\cite[sec. 6.578]{Gradshteyn:1943cpj}:
\ba
\int_0^\infty dx \;x^{\nu+1} K_{\mu} (a x)   J_{\mu}(b x) J_{\nu}(c x) &= \frac{(2c)^\nu (ab)^{\mu}}{(a^2-b^2+c^2)^{\mu+\nu+1}} \frac{\Gamma(\mu+\nu+1)}{\Gamma(\mu+1)}\\\
& \FF{\frac{\mu+\nu+1}{2}}{\frac{\mu+\nu+2}{2}}{\mu+1}{-\left(\frac{2ab}{a^2-b^2+c^2}\right)^2}~.
\ea
Finally, with the help of the hypergeometric identity
\be\label{eq:identity 2f1 sq to 2f1}
\FF{a}{b}{2b}{z}= \left(\frac{2}{2-z}\right)^a \FF{\frac{a}{2}}{\frac{a+1}{2}}{\frac{b+2}{2}}{\frac{z^2}{(2-z)^2}}~,
\ee
one finds the desired relation
\be
\text{FT}[\hat{D}_\D(\eta_2,\V{y}) K^{\pm}_\D(\eta_1,0,\V{y})]=\Pi_\D(\sigma)~.
\ee
\section{Large $\D$ limit of (A)dS propagators}\label{sec: large Delta}
In this section, we report the large $\D$ limit of the (A)dS propagators for a massive free scalar field. Before doing so, let us outline the widely-used Stirling’s approximation:
\be\label{eq:streling}
\lim_{\abs{z}\to\infty}\Gamma(z) = \sqrt{\frac{2\pi}{z}} e^{-z} z^z.
\ee
This approximation holds for any $z$ on the complex plane, except along the negative real line. In the main text, we encounter many Gamma functions with large imaginary parts. In other words, a Gamma function with $z=x+iy$ with $x \in \Real$, $y \in \Real^+$ and $y\gg1$. The Sterling's formula in this case takes the form:
\ba\label{eq: Stirling's}
\lim_{y\to\infty}\Gamma(x+iy) &= (1-i)\sqrt{\pi} e^{\frac{i \pi x}{2}-i y -\frac{\pi y}{2}} y^{i y + x - \half}~,\\
\lim_{y\to\infty}\Gamma(x-iy) &=\Gamma(x+iy)^*= (1+i)\sqrt{\pi} e^{-\frac{i \pi x}{2}+i y -\frac{\pi y}{2}} y^{-i y + x - \half}~,\\
\lim_{y\to\infty} \Gamma(x+ iy) \Gamma(x- iy)&=  2\pi y^{2x-1} e^{- \pi y}~.
\ea

Let us go back to the AdS and de Sitter propagators. They are the solutions of the Casimir/Laplacian equation in the respective spacetime. These propagators are normalized such that in short distance limit (also known as the coincident point limit) they reproduce the flat-space propagator. Their explicit expressions are given by
\ba
G_\D^\text{AdS} (\sigma)&= \frac{\Gamma(\D)}{2\pi^\hd\Gamma(\D-\hd+1)} \frac{1}{(-2-2\sigma)^\D} \FF{\D}{\D-\hd+\half}{2\D-d+1}{\frac{2}{1+\sigma}}~,\\
G_\D^\text{dS} (\sigma)&= \frac{\Gamma(\D)\Gamma(d-\D)}{(4\pi)^\frac{d+1}{2} \Gamma(\frac{d+1}{2})} \FF{\D}{d-\D}{\frac{d+1}{2}}{\frac{1+\sigma}{2}}~.
\ea
where the scaling dimension is related to the mass  with $\pm m^2R^2 = \D (d-\D)$, with minus sign for AdS and plus sign for de Sitter. Here, $\sigma$ is the chordal distance, defined as $\sigma = X_1\cdot X_2$ for AdS and $\sigma = Y_1\cdot Y_2$ for de Sitter. 

The large $\D$ limit can be derived from the integral representation of the hypergeometric functions 
\be\label{eq: 2F1 int rep}
\FF{a}{b}{c}{z}= \frac{1}{\Gamma(c-b)\Gamma(b)} \int_0^1 dx \;x^{b-1} (1-x)^{c-b-1} (1-z x )^{-a}~,
\ee 
and performing a saddle point approximation:
\ba\label{eq: large Delta lim}
 \lim_{|\D|\to\infty}G_\D^\text{AdS} (\sigma) &= \frac{e^{-i \pi  \Delta} \Delta ^{\frac{d}{2}-1}}{2^{d+1}\pi^\hd}  \frac{ (1-z)^{\Delta } }{z^{\frac{d}{4}}\left(\sqrt{z}+1\right)^{2 \Delta-d }}~,  \\
  \lim_{|\D|\to\infty} G_\D^\text{dS} (\sigma)&=\frac{ \Delta ^{\frac{d}{2}-1}}{2^{d+2} \pi^\hd} \frac{ (1-z)^{\Delta } }{z^{\frac{d}{4}}\left(\sqrt{z}+1\right)^{2 \Delta-d }}  \csc \left(\frac{ \pi}{2}  (d-2 \Delta )\right) ~+~ \D\leftrightarrow d-\D~,
\ea
where $z=\frac{\sigma+1}{\sigma-1}$.
 
\section{Contact terms}\label{sec: Contact terms}
In this section, we discuss the existence of contact terms (also called local terms) in the boundary theory of a free massive scalar field in de Sitter.
For a parallel discussion on the necessity of such boundary local terms from the viewpoint of unitarity, see~\cite{Hogervorst:2021uvp}. In addition, we will argue such terms generically exist in interacting theories. 

Consider the two-point function of a massive scalar field in de Sitter given by~\reef{eq: freeG}. In Fourier space, the two-point function is given by:
\be
\tilde{G}^{\text{free}}_\D(\V k) =  \frac{\pi}{4} (\eta_1\eta_2)^{\hd} H^{(1)}_{i\lambda}(-k\eta_1)H^{(2)}_{i\lambda}(-k\eta_2)~,
\ee
where we stripped out the momentum conservation delta function $(2\pi)^d \delta_{\V k, -\V k'}$ and $H^{(1)}, H^{(2)}$ are the Hankel functions of the first and second kind. Taking the late-time limit, we find
\ba
\tilde{G}^{\text{free}}_\D(\V k)\limu{\eta_1,\eta_2\to0^-} &\frac{1}{4\pi} \Bigg[(\eta_1 \eta_2)^\D\left( \left(\frac{k}{2}\right)^{-2i\lambda} \Gamma(i \lambda)^2 ~+~ c.c\right)\\
&\quad+(\eta_1 \eta_2)^\hd \left(e^{\pi \lambda} \Gamma(i\lambda)\Gamma(-i\lambda)  \left(\frac{\eta_1}{\eta_2}\right)^{i\lambda}~+~c.c\right)\Bigg]~.
\ea 
The first line above is responsible for creating the familiar power-law CFT two-point functions on the boundary, but unlike in AdS, because of the different boundary conditions in de Sitter, there exists the second line which is constant in momentum space and therefore creates the contact term on the boundary. In particular, by Fourier transforming back (for simplicity, we set $\eta_1=\eta_2=\eta$), one finds:
\ba\label{eq: late free G position}
{G}^{\text{free}}_\D(\sigma)\limu{\eta\to0^-} &(-\eta)^{2\D}\, \frac{\Gamma(\D)\Gamma(\hd-\D)}{4\pi^{\hd+1}}\,\frac{1}{y_{12}^{2\D}} ~+~ \D\longleftrightarrow \Db\\
&+(-\eta)^d\,\frac{\coth(\pi \lambda)}{2\lambda}\,\delta^{(d)}(\V y_1-\V y_2)~.
\ea 
Another way to see the existence of such contact terms is to look at the canonical commutation relation between bulk field $\phi$ and its conjugate $\Pi$:
\be\label{eq: canonical commutation}
[\phi(\eta,\V y_1), \Pi(\eta,\V y_2)] = (-\eta)^{1-d} [\phi(\eta,\V y_1), \partial_\eta\phi(\eta,\V y_2)] = i \delta_{\V y_1, \V y_2}~.
\ee
Now if one uses the bulk-to-boundary expansion for the free theory in~\reef{eq: free expansion}, one finds that the boundary operators of the free theory do not commute and, in particular, their commutator is given by a contact term:
\be
[\O_\D(\V y_1),\O_{\Db}(\V y_2)] = \frac{1}{2\lambda |\frak{g}_\D|}\delta_{\V y_1, \V y_2}~,
\ee
where we used the fact that $ |\frak{g}_\D| = \sqrt{ \frak{g}_\D \frak{g}_{\Db}} $.
One could also plug the bulk-to-boundary expansion~\reef{eq: free expansion} in~\reef{eq: late free G position}, find that  the de Sitter boundary two-point function has contact terms when the operators are shadow of each other and conclude:
\ba\label{eq: boundary two point free}
\Braket{\O_{\D}(\V y_1)\O_{\Db}(\V y_2)} &= \frac{\coth (\pi \lambda)+1}{4\lambda  |\frak{g}_\D|}\delta_{\V y_1, \V y_2}~,\\
 \Braket{\O_{\Db}(\V y_1)\O_{\D}(\V y_2)} &= \frac{\coth (\pi \lambda)-1}{4\lambda  |\frak{g}_\D|}\delta_{\V y_1, \V y_2}~.
\ea

So far we showed from both canonical quantization (the Lorentzain nature of de Sitter) and limit of the bulk two-point function, there exist contact terms (contact terms) at the boundary. Furthermore, we match the mentioned result to determine the normalization in~\reef{eq: boundary two point free}. Now, we will argue the existence of such terms in a generic theory. First, note that by using~\reef{eq: late free G position} and the \Kallen spectral decomposition, one can observe that such contact terms generally appear (to avoid clutter, here we again set the conformal times equal  $(\eta_1=\eta_2=\eta$)):
\be
\Braket{\phi(\eta,\V y_1)\phi(\eta,\V y_2)}\limu{\eta\to0^-}~\supset~  (-\eta)^d \delta^{(d)}(\V y_1-\V y_2) \int^\infty_{-\infty} \frac{d\lambda}{2\lambda}\,{\coth(\pi \lambda)} \,\rho(\D)~.
\ee
Note that the integrand above is non-negative leading to a non-zero integral for a generic spectral density (assuming the spectrum is non-trivial-- $\rho(\D) \neq0$).

\section{AdS harmonic analysis and de Sitter}\label{sec: Harmonic analysis}
Harmonic analysis in Euclidean AdS has been proven to be an exceptionally powerful tool. In this section, we briefly review key results from harmonic analysis in AdS before performing a Wick rotation to transition to de Sitter space. We denote the embedding space point in AdS by $X$ and the two-point invariant $\sigma^{\text{AdS}}=X_1\cdot X_2$. Harmonic functions in AdS are solutions of the Casimir equation with no singularity in short distances. These harmonic functions are given by:
\be
\Omega_\lambda(X_1,X_2)=\frac{1}{(4\pi)^\frac{d+1}{2} \Gamma(\frac{d+1}{2})} \frac{\Gamma(\hd\pm i\lambda)}{\Gamma(\pm i\lambda)}{}\FF{\hd+i\lambda}{\hd-i\lambda}{\frac{d+1}{2}}{\frac{1+\sigma^{\text{AdS}}}{2}}~.
\ee
They can also be written as sum of the bulk-to-bulk propagators since they are both solutions of the same Casimir equation: 
\be\label{eq: omega to pi}
\Omega_\lambda(X_1,X_2) = \frac{i\lambda}{2\pi}  \left( \Pi_{\hd+i\lambda}^{\text{AdS}}(\sigma^{\text{AdS}}) -\Pi_{\hd-i\lambda}^{\text{AdS}}(\sigma^{\text{AdS}}) \right)~.
\ee
Note that $\Pi_{\D}^{\text{AdS}}$  differs from $\Pi_{\D}$ in~\reef{eq:psi} by a normalization factor $\Pi_{\D}^{\text{AdS}}= \frac{\Gamma(\D)}{2\pi^{\hd}\Gamma(\D-\hd+1)}\Pi_{\D}$. 
Harmonic functions are orthogonal and satisfy  
\be
\int_{X} ~\Omega_{\lambda}(X_1,X) \Omega_{\lambda'}(X,X_2) = \half\left(\delta(\lambda+\lambda')+\delta(\lambda-\lambda')\right)\Omega_{\lambda'}(X_1,X_2)~,
\ee
where the integration is over AdS volume. For example, in Poincare coordinates it is given by:
\be
\int_{X}  = \int_0^\infty \frac{dz}{z^{d+1}} \int d^d\V x~.
\ee
The orthogonality relation above is equivalent to the completeness relation:
\be
\int_{-\infty}^\infty d\lambda~ \Omega_{\lambda}(X_1,X_2) = \delta(X_1,X_2)~.
\ee
Here,  $\delta(X_1,X_2)$ should be understood in a distributional sense in which the integration against an arbitrary function $\mathcal{F}(X_2)$ over AdS will return the same function at $X_1$:
\be
\int_{\text{AdS}} dX_2~\mathcal{F}(X_2) \delta(X_1,X_2)=  \mathcal{F}(X_1)~.
\ee
By substituting~\reef{eq: omega to pi} into the completeness relation above, one finds:
\be\label{eq:int over pi ads}
\int_{-\infty}^\infty \frac{i \lambda d\lambda}{\pi}~ \Pi^\text{AdS}_{\hd+i\lambda}(X_1,X_2) = \delta(X_1,X_2)~,
\ee
which turns out to be very useful shortly.

We may now perform a Wick rotation to de Sitter~\cite{Sleight:2021plv,DiPietro:2021sjt,Loparco:2023rug} in which we move from AdS in Poincare  coordinates to de Sitter planar coordinates by $z\to\pm i\eta $ and $\V x \to \V y$. This Wick rotation transforms $\sigma^{\text{AdS}} \to \sigma$ and therefore transforms the AdS harmonic function to free bulk two-point function:
\be
\Omega_\lambda(\sigma)\to \frac{1}{\Gamma(\pm i\lambda)}G_\D^{\text{free}}(\sigma)~,\qquad \Pi_{\D}^{\text{AdS}}(\sigma^{\text{AdS}})\to \frac{\Gamma(\D)}{2\pi^{\hd}\Gamma(\D-\hd+1)}\Pi_{\D}(\sigma)~.
\ee

Now, we would like to use the relations above to find a completeness relation similar to~\reef{eq:int over pi ads} for $\Pi_{\hd+i\lambda}$ where on the RHS we have a delta function in de Sitter. First notice $\delta(X_1,X_2)$ has only support for short distances in AdS  i.e. $\sigma^{\text{AdS}}=1$ while $\delta(Y_1,Y_2)$ corresponds to $\sigma^{\text{dS}}=-1$. So, one would like to find a completeness relation for $\Pi^\text{AdS}_{\hd+i\lambda}(X_1,-X_2)$\footnote{Strictly speaking $X_1$ and $X_2$ cannot be on the same AdS manifold. This should be seen in the sense of a complexified manifold.}. Fortunately, there is a straightforward relation between $\Pi_{\D}(\sigma)$ and $\Pi_{\D}(-\sigma)$. This relation follows from the hypergeometric identity:
\be
\FF{a}{b}{c}{z} = (1-z)^{-a}\FF{a}{c-b}{c}{\frac{z}{z-1}}~,
\ee
which is given by:
\be\label{psi -sigma}
\Pi_\D(-\sigma^\pm)= e^{\mp i\pi \D } \Pi_\D(\sigma^\pm)~.
\ee
Using this identity and applying the appropriate factors from the integration measure\footnote{The integral measures under Wick rotation transform as
\be
\int d^d\V x \int \frac{dz}{z^{d+1}} \to e^{\pm \frac{i \pi d}{2}}\int d^d\V y \int \frac{d\eta^\pm}{(-\eta^\pm)^{d+1}}~,
\ee
where $\eta^\pm = \eta\pm i \e$.
},
we find the completeness relation:
\be\label{eq: dS pi completenesss}
\int_{-\infty}^\infty d\lambda~e^{\pm \pi \lambda } \frac{\Gamma(\hd+i\lambda)}{2\pi^{\hd+1}\Gamma(i\lambda)}\Pi_{\D}(\sigma^\pm)= \delta(Y_1,Y_2)~.
\ee
One should exercise extra caution when using the equation above. The LHS is only a function of $\sigma$ and it does not differentiate two points on the light-cone from coincidental points. Therefore, the distribution above should be understood through the analytic continuation to AdS. Meaning that when performing an integral of the both sides of the equation above  against a test function, first two points will be analytically continued to AdS and then integration over spacetime will be performed in AdS and in the end one analytically continued back to de Sitter.
\section{de Sitter diagrams}\label{sec: Identites}
Here we list some identities that was used in the main text.
\subsection{The bubble diagram}\label{sec:bubblediagram}
The product of two-point functions in a CFT integrated over a common point:
\be\label{eq:bubble diagram}
\int \frac{d^d\V{y}}{|\V{y}_1-\V{y}|^{2\D}|\V{y}_2-\V{y}|^{2\Db}}= B_\D \delta_{\V{y}_1,\V{y}_2} \qquad \text{with} ~~ B_{\D} = \pi^d \frac{\Gamma(\hd-\D)\Gamma(\D-\hd)}{\Gamma(\D)\Gamma(d-\D)}
\ee
This identity is easily verifiable by performing a Fourier transform.
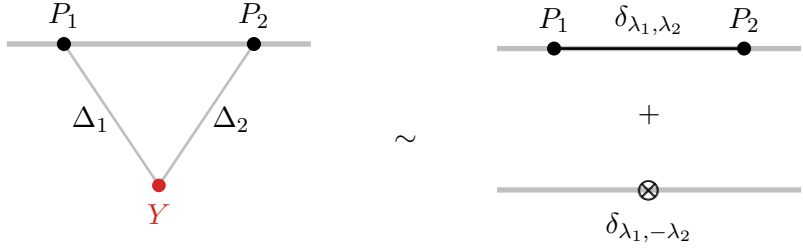
\begin{figure}[t!]
   \centering
           \begin{tabular}{cc}
\scalebox{1}{
\raisebox{-33pt}{
\begin{tikzpicture}[line width=1. pt, scale=2.5]
\draw[lightgray, line width=1.pt] (0,0.75) -- (0.5,0);
\draw[lightgray, line width=1.pt] (1,0.75) -- (0.5,0);
\draw[lightgray, line width=2.pt] (-0.3,0.75) -- (1.3,0.75);

\draw  (0.5,0) node[red3, fill, circle, scale=0.5]  {};
\draw[fill=black] (0,0.75) circle (.03cm);
\draw[fill=black] (1,0.75) circle (.03cm);

\node[scale=1] at (0,0.9) {$P_{1}$};
\node[scale=1] at (1,0.9) {$P_{2}$};
\node[red3,scale=1] at (0.5,-0.15){$Y$};
\node[scale=1] at (0.14,0.35){$\D_1$};
\node[scale=1] at (0.88,0.35){$\D_2$};

\end{tikzpicture}
}
}& \quad$\sim$\quad\quad
\scalebox{1}{
\raisebox{-40pt}{
\begin{tikzpicture}[line width=1. pt, scale=2.5]
\draw[lightgray, line width=2.pt] (-0.3,0.75) -- (1.3,0.75);
\draw[black, line width=1.pt] (0,0.75) -- (1,0.75);

\draw[fill=black] (0,0.75) circle (.03cm);
\draw[fill=black] (1,0.75) circle (.03cm);

\node[scale=1] at (0.5,0.9) {$\delta_{\lambda_1,\lambda_2}$};
\node[scale=1] at (0,0.9) {$P_{1}$};
\node[scale=1] at (1,0.9) {$P_{2}$};

\node[scale=1] at (0.5,0.4) {$+$};
\draw[lightgray, line width=2.pt] (-0.3,0) -- (1.3,0);
\node[scale=1] at (0.5,0) {$\pmb{\otimes}$};
\node[scale=1] at (0.5,-0.2) {$\delta_{\lambda_1,-\lambda_2}$};
\end{tikzpicture}
}
}
\end{tabular}
   \caption{The V diagram~\reef{eq:Vdiag}: The product of two bulk-to-boundary operators integrated over the bulk point produces the standard CFT two-point function as well as a contact term.}
  \label{fig:V diagram }
\end{figure}
\subsection{The V diagram}\label{sec: V diagram}
The integral over the  common bulk point of two bulk-to-boundary propagator~\cite{Laddha:2022nmj} is given by: 
\ba\label{eq:Vdiag}
\int \frac{d^d\V y d\eta}{(-\eta)^{d+1}} K^{\pm}_{\D_1}(\eta,\V{y};\V{y}_1) K^{\pm}_{\D_2}(\eta,\V{y};\V{y}_2)= \frac{2\pi^{\hd+1} \Gamma(i\lambda_1)}{e^{\pm\pi \lambda_1}\Gamma(\hd+i\lambda_1)}   H_{\D_1,\D_2} (\V{y}_1,\V{y}_2)~,
\ea
where the bulk-to-boundary propagators have the same $i\e$ prescription and 
\be\label{eq: H}
H^\pm_{\D_1,\D_2} (\V{y}_1,\V{y}_2) \equiv \frac{1}{y_{12}^{2\D_1}} \delta_{\lambda_1 \lambda_2} + \xi^\pm_{\Db_1} \delta_{\lambda_1,-\lambda_2} \delta_{\V{y}_1,\V{y}_2}~, \quad \text{with}~ \xi^\pm_\D = \pi^\hd  \frac{\Gamma( i\lambda)}{e^{\pm\pi\lambda}\Gamma(\hd+i\lambda)}~.
\ee
One could derive the above expression by going to Fourier space~\cite{Laddha:2022nmj}. Using the equation~\reef{eq: FT of Kpm} one can rewrite the V diagram as:
\ba
\int& \frac{d^d\V y d\eta}{(-\eta)^{d+1}} K^{\pm}_{\D_1}(\eta,\V{y};\V{y}_1) K^{\pm}_{\D_2}(\eta,\V{y};\V{y}_2)=- \int_\eta  \frac{\pi^{d+2}  2^{-i(\lambda_1+\lambda_2)} e^{\mp\pi(  \lambda_1+ \lambda_2 )} (-\eta)^d }{ (2\pi)^{2d}  \Gamma(\hd+i\lambda_1)\Gamma(\hd+i\lambda_2)} \times\\
 &\times\int_{\V y , \V k_1 \V k_2}   k_1^{i\lambda_1} k_2^{i\lambda_2} e^{i \V k_1\cdot(\V y - \V y_1)}e^{i \V k_2\cdot(\V y - \V y_2)} H^{(1)/(2)}_{i\lambda_1}(-k_1\eta)H^{(1)/(2)}_{i\lambda_2}(-k_2\eta)~.
\ea
After integrating over $\V y$, which produces a Dirac delta function, and performing the integral over one of the spatial momentums, we find:
\ba
\int& \frac{d^d\V y d\eta}{(-\eta)^{d+1}} K^{\pm}_{\D_1}(\eta,\V{y};\V{y}_1) K^{\pm}_{\D_2}(\eta,\V{y};\V{y}_2)= - \frac{\pi^{d+2}  2^{-i(\lambda_1+\lambda_2)}  e^{\mp\pi(  \lambda_1+ \lambda_2 )} }{ (2\pi)^{d}  \Gamma(\hd+i\lambda_1)\Gamma(\hd+i\lambda_2)} \times\\
 &\times\int_{ \V k}   k^{i(\lambda_1+\lambda_2)} e^{i \V k\cdot(\V y_1 - \V y_2)} \int^0_{-\infty} \frac{d\eta}{\eta} H^{(1)/(2)}_{i\lambda_1}(-k\eta)H^{(1)/(2)}_{i\lambda_2}(-k\eta)~.
\ea
The second integral above can be extracted from the orthogonality relation of the Kontorovich–Lebedev transform found in~\reef{eq: Kontorovich–Lebedev}  as well as the relation between Hankel and Bessel function of the first kind. More precisely:
\be
\int_0^\infty \frac{dx}{x}   H^{(1)/(2)}_{i\lambda_1}(x)H^{(1)/(2)}_{i\lambda_2}(x) = \frac{-2}{\lambda_1\sinh \pi \lambda_1}\left(\delta_{\lambda_1,-\lambda_2}+e^{\pm \pi \lambda}\delta_{\lambda_1,\lambda_2}\right)~.
\ee
Again, the notation above means that the plus sign is for the integral of the product of two $H^{(1)}$ functions, and the minus corresponds to the integral of the product of two $H^{(2)}$ functions.
Using this relation, the integral above splits into two parts: one is the Fourier transform of a constant and the other is a Fourier transform of a monomial given by~\reef{eq FT of power}. In the end, one finds that the V diagram is given by:
\ba\label{eq: KK V diagram}
\int& \frac{d^d\V y d\eta}{(-\eta)^{d+1}} K^{\pm}_{\D_1}(\eta,\V{y};\V{y}_1) K^{\pm}_{\D_2}(\eta,\V{y};\V{y}_2)=\\ 
 &\qquad \qquad\frac{2\pi^{d+1} \Gamma( i\lambda_1)\Gamma(- i\lambda_1)}{  \Gamma(\hd+ i\lambda_1)\Gamma(\hd- i\lambda_1)} \delta_{\lambda_1,-\lambda_2}\delta_{\V y_1, \V y_2} + 
\frac{2\pi^{\hd+1}  \Gamma( i\lambda_1) }{  e^{\pm\pi\lambda_1 }  \Gamma(\hd+i\lambda_1)}\frac{\delta_{\lambda_1,\lambda_2}}{y_{12}^{2\D_1}}~.
\ea
Now with reordering and factoring the coefficient of the CFT two-point function, one finds the equation~\reef{eq:bubble diagram}. 

\subsection{de Sitter split representation}\label{sec: Split rep}
In this section, we derive a similar relation to AdS split representation~\cite{Costa:2014kfa} for de sitter, in which the product of two bulk-to-boundary propagators with shadow dimensions integrated over a boundary point is proportional to the bulk-to-bulk propagator~\cite{Sleight:2019mgd,Sleight:2019hfp,Loparco:2023rug}:
\ba\label{eq: ds split rep}
\int d^d\V{y}~ K^\pm_{\D}(\eta_1,\V{y}_1;\V{y}) K^\mp_{\Db}(\eta_2,\V{y}_2;\V{y})&=   {e^{\mp\pi \lambda}} \frac{\pi^\frac{d+1}{2}}{2^{d-1}\Gamma(\frac{d+1}{2})}\FF{\D}{d-\D}{\frac{d+1}{2}}{\frac{1+\sigma}{2}}\\
&= {e^{\mp\pi \lambda}} \frac{4 \pi^{d+1}} {\Gamma(\D)\Gamma(d-\D)}G_\D^{\text{free}}(\sigma)
\ea
where the bulk-to-boundary propagators have opposite $i\e$ prescriptions. 
One way to prove this identity is directly going to Fourier space. Here, we take a different approach.  We prove this relation by rewriting the bulk-to-boundary propagator using the boundary-to-bulk connector. This approach highlights the remarkable power of the boundary-to-bulk connector in de Sitter calculations. Substituting~\reef{eq:DtoBulkBoundary} on the LHS of the relation above, it produces four terms:
\ba
&\Dh{}{1}\Dhb{}{2} \int_{\V y}\frac{1}{|\V {y}-\V{y}_1|^{2\D}}\frac{1}{|\V {y}-\V{y}_2|^{2\Db}}\,+\,\xi^{\pm}_{\D}\xi^{\mp}_{\Db}\, \Dhb{}{1}\Dh{}{2}\, \int_{\V y} \delta_{\V{y},\V{y}_1}  \delta_{\V{y},\V{y}_2} \\
&+\xi^{\mp}_{\Db}\Dh{}{1}\Dh{}{2} \int_{\V y}\frac{\delta_{\V y, \V y_2}}{|\V {y}-\V{y}_1|^{2\D}}+\xi^{\pm}_{\D}\Dhb{}{1}\Dhb{}{2} \int_{\V y}\frac{\delta_{\V y, \V y_1}}{|\V {y}-\V{y}_2|^{2\Db}}~.
\ea
The last three integrals are trivial and the first integral corresponds to the bubble diagram in equation~\reef{eq:bubble diagram}:
\ba
&B_\D \Dh{}{1}\Dhb{}{2} \delta_{\V{y}_1,\V{y}_2} \,+\,\xi^{\pm}_{\D}\xi^{\mp}_{\Db}\, \Dhb{}{1}\Dh{}{2}\, \delta_{\V{y}_1,\V{y}_2}  \\
&+\xi^{\mp}_{\Db}\Dh{}{1}\Dh{}{2} \frac{1}{y_{12}^{2\D}}+\xi^{\pm}_{\D}\Dhb{}{1}\Dhb{}{2} \frac{1}{y_{12}^{2\Db}}~,
\ea
where we used the distributional identity that $ \int_{\V y} \delta_{\V{y},\V{y}_1}  \delta_{\V{y},\V{y}_2} =  \delta_{\V{y}_1,\V{y}_2}$. The first and the third term as well as the second and the fourth term can be grouped by factoring out $\Dh{}{1}$ and $\Dhb{}{1}$ respectively:
\ba
&\xi^{\mp}_{\Db}\Dh{}{1}\left[\Dh{}{2} \frac{1}{y_{12}^{2\D}}+\xi^{\mp}_{\D} \Dhb{}{2}\ \delta_{\V{y}_1,\V{y}_2}\right] \\
&+\xi^{\pm}_{\D}\, \Dhb{}{1}\left[\Dhb{}{2} \frac{1}{y_{12}^{2\Db}} + \xi^{\mp}_{\Db} \Dh{}{2}\, \delta_{\V{y}_1,\V{y}_2}\right]~,
\ea
where we used the fact that $B_\D=\xi^{\pm}_{\D}\xi^{\pm}_{\Db}$. Each square bracket above makes the bulk-to-boundary propagator according to~\reef{eq:DtoBulkBoundary}.
Now, using~\reef{eq: psi from D on K } one finds:
\be
\xi^{\mp}_{\Db}\Pi_\D(\sigma)+\xi^{\pm}_{\D}\,\Pi_{\Db}(\sigma)=  {e^{\mp\pi \lambda}}\frac{\pi^\frac{d+1}{2}}{2^{d-1}\Gamma(\frac{d+1}{2})}\FF{\D}{\Db}{\frac{d+1}{2}}{\frac{1+\sigma}{2}}~,
\ee
which considering the relation~\reef{eq:psi + psi G} yields the desired split representation identity.
\begin{figure}[t!]
   \centering
           \begin{tabular}{cc}
\scalebox{1}{
\raisebox{-33pt}{
\begin{tikzpicture}[line width=1. pt, scale=2.5]
\draw[lightgray, line width=1.pt] (0,0) -- (0.5,0.75);
\draw[lightgray, line width=1.pt] (1,0) -- (0.5,0.75);
\draw[lightgray, line width=2.pt] (-0.3,0.75) -- (1.3,0.75);
\draw[fill=black] (0,0) circle (.03cm);
\draw[fill=black] (1,0) circle (.03cm);
\node[scale=1] at (0,-.15) {$Y_{1}$};
\node[scale=1] at (1,-.15) {$Y_{2}$};
\draw  (0.5,0.75) node[blue3, fill, circle, scale=0.5]  {};
\node[blue3,scale=1] at (0.5,0.9){$P$};
\node[scale=1] at (0.1,0.38){$\D$};
\node[scale=1] at (0.9,0.38){$\Db$};

\end{tikzpicture}
}
}& ~~$\propto$~~
\scalebox{1}{
\raisebox{-33pt}{
\begin{tikzpicture}[line width=1. pt, scale=2.5]
\draw[fill=black] (0,0) -- (1,0);
\draw[lightgray, line width=2.pt] (-0.3,0.75) -- (1.3,0.75);
\draw[fill=black] (0,0) circle (.03cm);
\draw[fill=black] (1,0) circle (.03cm);
\node[scale=1] at (0,-.15) {$Y_{1}$};
\node[scale=1] at (1,-.15) {$Y_{2}$};

\end{tikzpicture}
}
}
\end{tabular}
   \caption{The split representation in de Sitter~\reef{eq: ds split rep}: The representation of the free bulk-to-bulk two-point function in terms of two bulk-to-boundary propagators with shadow dimensions integrated over the boundary point.}
  \label{fig:Split representation}
\end{figure}
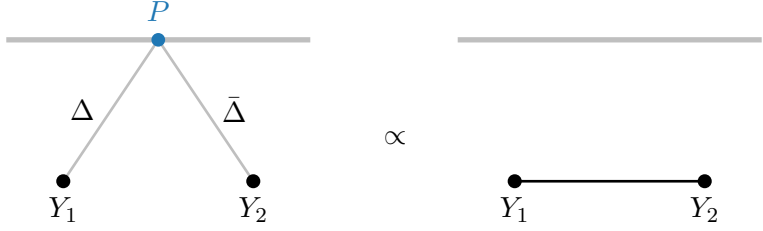

\subsection{The broken leg}\label{sec: broken leg}
In multiple instances in the main text, we encounter the \textit{broken leg} diagram, which is defined as the following integral over a  bulk point shared by the product of a bulk-to-boundary propagator and  a bulk-to-bulk two-point function:
\be
\frak{B}=\int \frac{d^dy d\eta}{(-\eta)^{d+1}} K^\pm_{\D_1}(\eta,\V{y};\V{y}_1) G^\text{free}_{\D_2} (\eta,\V{y};\eta_2,\V{y}_2)~.
\ee
We will derive an explicit formula for this diagram. First, notice that one can find out that the answer should be proportional to the bulk-to-boundary propagator based on a scaling and dS invariance argument similar to the inversion formula heuristic derivation. Here, we will perform the integral explicitly to find the proportionality coefficient. We start with the split representation~\reef{eq: ds split rep} and rewrite the  bulk-to-bulk two-point function:
\be
\frak{B}= e^{\pm\pi \lambda}\frac {\Gamma(\D)\Gamma(d-\D)}{4 \pi^{d+1}}\int \frac{d^d\V y d^d\V y' d\eta}{(-\eta)^{d+1}} K^\pm_{\D_1}(\eta,\V{y};\V{y}_1) K^\pm_{\D_2}(\eta,\V{y};\V{y}') K^\mp_{\Db_2}(\eta_2,\V{y}_2;\V{y}')~.
\ee
The integral over the bulk point $Y=(\eta,\V y)$ corresponds to the V diagram in equation~\reef{eq:Vdiag}:
\be
\frak{B}= \frac {\Gamma(d-\D_2)\Gamma(\D_2-\hd)}{2\pi^{\hd}} \int {d^d\V y'}   H_{\D_2,\D_1} (\V{y}_1,\V{y}') K^\mp_{\Db_2}(\eta_2,\V{y}_2;\V{y}')~.
\ee
The function $H$ includes two parts as shown in~\reef{eq: H}. The integration over the  Dirac delta part  is trivial and it indeed produces a term proportional to the bulk-to-boundary propagator:
\be
\frak{B} \supset \frac{\Gamma( i\lambda_1)\Gamma(- i\lambda_1)}{2e^{\pm\pi\lambda_1}}\delta_{\lambda_1,-\lambda_2} K^\mp_{\D_1}(\eta_2,\V{y}_2;\V{y}_1)~.
\ee
 To find the other part  of the integral above, we use the power of boundary-to-bulk connector:
 \ba
 \frak{B} \supset& \frac {\Gamma(\hd-i\lambda_2)\Gamma(i\lambda_2)}{2\pi^{\hd}} \delta_{\lambda_1,\lambda_2}\left[\hat{D}_{\Db_1}\int_{\V y'} \frac{1}{|\V y_1 - \V y'|^{2\D_1}} \frac{1}{|\V y_2- \V y'|^{2\Db_1}}+\xi^\mp_{\Db_1} \hat{D}_{\D_1} \int_{\V y'} \frac{\delta_{\V y_2,\V y'}}{|\V y_1- \V y'|^{2\D_1}}\right]\\
 &= \delta_{\lambda_1,\lambda_2}  \frac{\Gamma( i \lambda_1)\Gamma(- i \lambda_1)}{2 e^{\pm\pi\lambda_1}}  \left[ \hat{D}_{\D_1}  \frac{1}{y_{12}^{2\D_1}} +  \xi^{\mp}_{\D_1}\hat{D}_{\Db_1} \delta_{\V y_1 , \V y_2} \right] \\
 &= \frac{\Gamma( i\lambda_1)\Gamma(- i\lambda_1)}{2e^{\pm\pi\lambda_1}}\delta_{\lambda_1,\lambda_2} K^\mp_{\D_1}(\eta_2,\V{y}_2;\V{y}_1)~.
 \ea 
Combining both terms, we find that the broken leg diagram is given by:
 \ba\label{eq:broken leg id}
 \int& \frac{d^dy d\eta}{(-\eta)^{d+1}} K^\pm_{\D_1}(\eta,\V{y};\V{y}_1)  G^\text{free}_{\D_2} (\eta,\V{y};\eta_2,\V{y}_2)\\
&\qquad\qquad\qquad =  \frac{\Gamma( i\lambda_1)\Gamma(- i\lambda_1)}{2e^{\pm\pi\lambda_1}}K^\mp_{\D_1}(\eta_2,\V{y}_2;\V{y}_1)\left(\delta_{\lambda_1,\lambda_2} +\delta_{\lambda_1,-\lambda_2} \right)~.
 \ea
\begin{figure}[t!]
   \centering
           \begin{tabular}{cc}
\scalebox{1}{
\raisebox{-33pt}{
\begin{tikzpicture}[line width=1. pt, scale=2.5]
\draw[lightgray, line width=1.pt]  (0,0.25)  -- (0.5,0.75);
\draw[lightgray, line width=2.pt] (-0.3,0.75) -- (1.3,0.75);

\draw  (0.5,0.75) node[fill, circle, scale=0.5]  {};
\draw[fill=black] (0,0.25) -- (1,0);

\draw  (0,0.25)  node[red3,fill, circle, scale=0.5]  {};
\draw[fill=black] (1,0) circle (.03cm);

\node[scale=1] at (0.5,0.9){$P$};
\node[scale=1] at (0.12,0.55){$\D_1$};
\node[scale=1] at (0.6,0.22){$\D_2$};
\node[scale=1,red3] at (0,0.1) {$Y'$};
\node[scale=1] at (1,-.15) {$Y$};
\end{tikzpicture}
}
}& ~~$\propto$~~ $ \left(\delta_{\lambda_1,\lambda_2}+\delta_{\lambda_1,-\lambda_2}\right) ~\times$ 
\scalebox{1}{
\raisebox{-33pt}{
\begin{tikzpicture}[line width=1. pt, scale=2.5]

\draw[lightgray, line width=1.pt] (1,0) -- (0.5,0.75);
\draw[lightgray, line width=2.pt] (0.3,0.75) -- (1.3,0.75);

\draw  (0.5,0.75) node[fill, circle, scale=0.5]  {};
\draw[fill=black] (1,0) circle (.03cm);

\node[scale=1] at (0.5,0.9){$P$};
\node[scale=1] at (0.65,0.3){$\D_1$};
\node[scale=1] at (1,-.15) {$Y$};
\end{tikzpicture}
}
}

\end{tabular}
   \caption{The broken leg diagram~\reef{eq:broken leg id}: The product of a free bulk two-point function and a bulk-to-boundary propagator integrated over the bulk point is proportional to a bulk-to-boundary propagator with Dirac delta of their scaling dimensions.}
  \label{fig:broken leg diagram}
\end{figure}
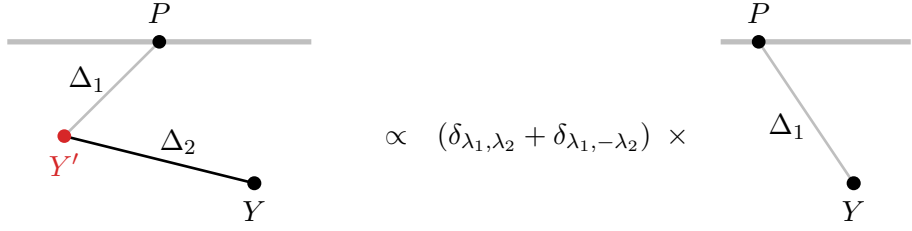
\bibliographystyle{JHEP}
\bibliography{bibliography}
\end{document}